\documentclass[longauth]{aa}  
\usepackage{graphicx}
\usepackage{txfonts}
\usepackage{dcolumn}
\usepackage{braket}
\usepackage{tabularx,rotating,multirow}
\usepackage{float}
\usepackage[caption=false]{subfig}
\usepackage{hyperref}
\usepackage{amsfonts}

\usepackage[nameinlink]{cleveref}

\makeatletter
\renewcommand*\aa@pageof{, page \thepage{} of \pageref*{LastPage}}
\makeatother

\begin{document} 

   \title{\emph{Euclid} preparation: XXVIII. Modelling of the weak lensing angular power spectrum}

%% please do not edit the author list -- contact ECEB Bureau for changes
\newcommand{\orcid}[1]{} %% define as link to https://orcid.org/#1 if needed
\author{Euclid Collaboration: A.~C.~Deshpande\orcid{0000-0003-3721-4232}$^{1}$, T.~Kitching\orcid{0000-0002-4061-4598}$^{1}$\thanks{\email{t.kitching@ucl.ac.uk}}, A.~Hall\orcid{0000-0002-3139-8651}$^{2}$, M.~L.~Brown\orcid{0000-0002-0370-8077}$^{3}$, N.~Aghanim$^{4}$, L.~Amendola$^{5}$, N.~Auricchio\orcid{0000-0003-4444-8651}$^{6}$, M.~Baldi\orcid{0000-0003-4145-1943}$^{7,6,8}$, R.~Bender\orcid{0000-0001-7179-0626}$^{9,10}$, D.~Bonino$^{11}$
% ...
}
   
%% please do not edit the affiliation list -- contact ECEB Bureau for changes
%% please do not edit the affiliation list -- contact ECEB Bureau for changes
\institute{$^{1}$ Mullard Space Science Laboratory, University College London, Holmbury St Mary, Dorking, Surrey RH5 6NT, UK\\
$^{2}$ Institute for Astronomy, University of Edinburgh, Royal Observatory, Blackford Hill, Edinburgh EH9 3HJ, UK\\
$^{3}$ Jodrell Bank Centre for Astrophysics, Department of Physics and Astronomy, University of Manchester, Oxford Road, Manchester M13 9PL, UK\\
$^{4}$ Universit\'e Paris-Saclay, CNRS, Institut d'astrophysique spatiale, 91405, Orsay, France\\
$^{5}$ Institut f\"ur Theoretische Physik, University of Heidelberg, Philosophenweg 16, 69120 Heidelberg, Germany\\
$^{6}$ INAF-Osservatorio di Astrofisica e Scienza dello Spazio di Bologna, Via Piero Gobetti 93/3, 40129 Bologna, Italy\\
$^{7}$ Dipartimento di Fisica e Astronomia "Augusto Righi" - Alma Mater Studiorum Universit\`{a} di Bologna, via Piero Gobetti 93/2, 40129 Bologna, Italy\\
$^{8}$ INFN-Sezione di Bologna, Viale Berti Pichat 6/2, 40127 Bologna, Italy\\
$^{9}$ Max Planck Institute for Extraterrestrial Physics, Giessenbachstr. 1, 85748 Garching, Germany\\
$^{10}$ Universit\"ats-Sternwarte M\"unchen, Fakult\"at f\"ur Physik, Ludwig-Maximilians-Universit\"at M\"unchen, Scheinerstrasse 1, 81679 M\"unchen, Germany\\
$^{11}$ INAF-Osservatorio Astrofisico di Torino, Via Osservatorio 20, 10025 Pino Torinese (TO), Italy\\
$^{12}$ Dipartimento di Fisica, Universit\`{a} di Genova, Via Dodecaneso 33, 16146, Genova, Italy\\
$^{13}$ INFN-Sezione di Roma Tre, Via della Vasca Navale 84, 00146, Roma, Italy\\
$^{14}$ Department of Physics "E. Pancini", University Federico II, Via Cinthia 6, 80126, Napoli, Italy\\
$^{15}$ Instituto de Astrof\'isica e Ci\^encias do Espa\c{c}o, Universidade do Porto, CAUP, Rua das Estrelas, PT4150-762 Porto, Portugal\\
$^{16}$ Dipartimento di Fisica, Universit\'a degli Studi di Torino, Via P. Giuria 1, 10125 Torino, Italy\\
$^{17}$ INFN-Sezione di Torino, Via P. Giuria 1, 10125 Torino, Italy\\
$^{18}$ INAF-IASF Milano, Via Alfonso Corti 12, 20133 Milano, Italy\\
$^{19}$ INAF-Osservatorio Astronomico di Roma, Via Frascati 33, 00078 Monteporzio Catone, Italy\\
$^{20}$ INFN-Sezione di Roma, Piazzale Aldo Moro, 2 - c/o Dipartimento di Fisica, Edificio G. Marconi, 00185 Roma, Italy\\
$^{21}$ Institut de F\'{i}sica d'Altes Energies (IFAE), The Barcelona Institute of Science and Technology, Campus UAB, 08193 Bellaterra (Barcelona), Spain\\
$^{22}$ Port d'Informaci\'{o} Cient\'{i}fica, Campus UAB, C. Albareda s/n, 08193 Bellaterra (Barcelona), Spain\\
$^{23}$ Institut d'Estudis Espacials de Catalunya (IEEC), Carrer Gran Capit\'a 2-4, 08034 Barcelona, Spain\\
$^{24}$ Institute of Space Sciences (ICE, CSIC), Campus UAB, Carrer de Can Magrans, s/n, 08193 Barcelona, Spain\\
$^{25}$ INAF-Osservatorio Astronomico di Capodimonte, Via Moiariello 16, 80131 Napoli, Italy\\
$^{26}$ INFN section of Naples, Via Cinthia 6, 80126, Napoli, Italy\\
$^{27}$ Dipartimento di Fisica e Astronomia "Augusto Righi" - Alma Mater Studiorum Universit\'a di Bologna, Viale Berti Pichat 6/2, 40127 Bologna, Italy\\
$^{28}$ Centre National d'Etudes Spatiales, Toulouse, France\\
$^{29}$ Institut national de physique nucl\'eaire et de physique des particules, 3 rue Michel-Ange, 75794 Paris C\'edex 16, France\\
$^{30}$ ESAC/ESA, Camino Bajo del Castillo, s/n., Urb. Villafranca del Castillo, 28692 Villanueva de la Ca\~nada, Madrid, Spain\\
$^{31}$ European Space Agency/ESRIN, Largo Galileo Galilei 1, 00044 Frascati, Roma, Italy\\
$^{32}$ Institute of Physics, Laboratory of Astrophysics, Ecole Polytechnique F\'{e}d\'{e}rale de Lausanne (EPFL), Observatoire de Sauverny, 1290 Versoix, Switzerland\\
$^{33}$ Departamento de F\'isica, Faculdade de Ci\^encias, Universidade de Lisboa, Edif\'icio C8, Campo Grande, PT1749-016 Lisboa, Portugal\\
$^{34}$ Instituto de Astrof\'isica e Ci\^encias do Espa\c{c}o, Faculdade de Ci\^encias, Universidade de Lisboa, Campo Grande, 1749-016 Lisboa, Portugal\\
$^{35}$ Department of Astronomy, University of Geneva, ch. d'Ecogia 16, 1290 Versoix, Switzerland\\
$^{36}$ Department of Physics, Oxford University, Keble Road, Oxford OX1 3RH, UK\\
$^{37}$ Universit\'e Paris-Saclay, Universit\'e Paris Cit\'e, CEA, CNRS, Astrophysique, Instrumentation et Mod\'elisation Paris-Saclay, 91191 Gif-sur-Yvette, France\\
$^{38}$ Univ Lyon, Univ Claude Bernard Lyon 1, CNRS/IN2P3, IP2I Lyon, UMR 5822, 69622, Villeurbanne, France\\
$^{39}$ INAF-Osservatorio Astronomico di Trieste, Via G. B. Tiepolo 11, 34143 Trieste, Italy\\
$^{40}$ INAF-Osservatorio Astronomico di Padova, Via dell'Osservatorio 5, 35122 Padova, Italy\\
$^{41}$ Institute of Theoretical Astrophysics, University of Oslo, P.O. Box 1029 Blindern, 0315 Oslo, Norway\\
$^{42}$ Leiden Observatory, Leiden University, Niels Bohrweg 2, 2333 CA Leiden, The Netherlands\\
$^{43}$ Jet Propulsion Laboratory, California Institute of Technology, 4800 Oak Grove Drive, Pasadena, CA, 91109, USA\\
$^{44}$ Technical University of Denmark, Elektrovej 327, 2800 Kgs. Lyngby, Denmark\\
$^{45}$ Cosmic Dawn Center (DAWN), Denmark\\
$^{46}$ Institut d'Astrophysique de Paris, 98bis Boulevard Arago, 75014, Paris, France\\
$^{47}$ Max-Planck-Institut f\"ur Astronomie, K\"onigstuhl 17, 69117 Heidelberg, Germany\\
$^{48}$ Aix-Marseille Universit\'e, CNRS/IN2P3, CPPM, Marseille, France\\
$^{49}$ Universit\'e de Gen\`eve, D\'epartement de Physique Th\'eorique and Centre for Astroparticle Physics, 24 quai Ernest-Ansermet, CH-1211 Gen\`eve 4, Switzerland\\
$^{50}$ Department of Physics, P.O. Box 64, 00014 University of Helsinki, Finland\\
$^{51}$ Helsinki Institute of Physics, Gustaf H{\"a}llstr{\"o}min katu 2, University of Helsinki, Helsinki, Finland\\
$^{52}$ NOVA optical infrared instrumentation group at ASTRON, Oude Hoogeveensedijk 4, 7991PD, Dwingeloo, The Netherlands\\
$^{53}$ Argelander-Institut f\"ur Astronomie, Universit\"at Bonn, Auf dem H\"ugel 71, 53121 Bonn, Germany\\
$^{54}$ Department of Physics, Institute for Computational Cosmology, Durham University, South Road, DH1 3LE, UK\\
$^{55}$  Universit\'e Paris Cit\'e, CNRS, Astroparticule et Cosmologie, 75013 Paris, France\\
$^{56}$ Institut d'Astrophysique de Paris, UMR 7095, CNRS, and Sorbonne Universit\'e, 98 bis boulevard Arago, 75014 Paris, France\\
$^{57}$ European Space Agency/ESTEC, Keplerlaan 1, 2201 AZ Noordwijk, The Netherlands\\
$^{58}$ Kapteyn Astronomical Institute, University of Groningen, PO Box 800, 9700 AV Groningen, The Netherlands\\
$^{59}$ Department of Physics and Astronomy, University of Aarhus, Ny Munkegade 120, DK-8000 Aarhus C, Denmark\\
$^{60}$ AIM, CEA, CNRS, Universit\'{e} Paris-Saclay, Universit\'{e} de Paris, 91191 Gif-sur-Yvette, France\\
$^{61}$ Space Science Data Center, Italian Space Agency, via del Politecnico snc, 00133 Roma, Italy\\
$^{62}$ Institute of Space Science, Bucharest, 077125, Romania\\
$^{63}$ Dipartimento di Fisica e Astronomia "G.Galilei", Universit\'a di Padova, Via Marzolo 8, 35131 Padova, Italy\\
$^{64}$ INFN-Padova, Via Marzolo 8, 35131 Padova, Italy\\
$^{65}$ Dipartimento di Fisica e Astronomia, Universit\'a di Bologna, Via Gobetti 93/2, 40129 Bologna, Italy\\
$^{66}$ Departamento de F\'isica, FCFM, Universidad de Chile, Blanco Encalada 2008, Santiago, Chile\\
$^{67}$ Institut f\"ur Astro- und Teilchenphysik, Universit\"at Innsbruck, Technikerstr. 25/8, 6020 Innsbruck, Austria\\
$^{68}$ Institut de Ciencies de l'Espai (IEEC-CSIC), Campus UAB, Carrer de Can Magrans, s/n Cerdanyola del Vall\'es, 08193 Barcelona, Spain\\
$^{69}$ Centro de Investigaciones Energ\'eticas, Medioambientales y Tecnol\'ogicas (CIEMAT), Avenida Complutense 40, 28040 Madrid, Spain\\
$^{70}$ Instituto de Astrof\'isica e Ci\^encias do Espa\c{c}o, Faculdade de Ci\^encias, Universidade de Lisboa, Tapada da Ajuda, 1349-018 Lisboa, Portugal\\
$^{71}$ Universidad Polit\'ecnica de Cartagena, Departamento de Electr\'onica y Tecnolog\'ia de Computadoras, 30202 Cartagena, Spain\\
$^{72}$ Institut de Recherche en Astrophysique et Plan\'etologie (IRAP), Universit\'e de Toulouse, CNRS, UPS, CNES, 14 Av. Edouard Belin, 31400 Toulouse, France\\
$^{73}$ Infrared Processing and Analysis Center, California Institute of Technology, Pasadena, CA 91125, USA\\
$^{74}$ IFPU, Institute for Fundamental Physics of the Universe, via Beirut 2, 34151 Trieste, Italy\\
$^{75}$ INAF-Osservatorio Astronomico di Brera, Via Brera 28, 20122 Milano, Italy\\
$^{76}$ Instituto de Astrof\'isica de Canarias, Calle V\'ia L\'actea s/n, 38204, San Crist\'obal de La Laguna, Tenerife, Spain\\
$^{77}$ Center for Computational Astrophysics, Flatiron Institute, 162 5th Avenue, 10010, New York, NY, USA\\
$^{78}$ School of Physics and Astronomy, Cardiff University, The Parade, Cardiff, CF24 3AA, UK\\
$^{79}$ INAF-Istituto di Astrofisica e Planetologia Spaziali, via del Fosso del Cavaliere, 100, 00100 Roma, Italy\\
$^{80}$ Department of Physics and Helsinki Institute of Physics, Gustaf H\"allstr\"omin katu 2, 00014 University of Helsinki, Finland\\
$^{81}$ Junia, EPA department, 59000 Lille, France\\
$^{82}$ INFN-Bologna, Via Irnerio 46, 40126 Bologna, Italy\\
$^{83}$ Instituto de F\'isica Te\'orica UAM-CSIC, Campus de Cantoblanco, 28049 Madrid, Spain\\
$^{84}$ CERCA/ISO, Department of Physics, Case Western Reserve University, 10900 Euclid Avenue, Cleveland, OH 44106, USA\\
$^{85}$ Laboratoire de Physique de l'\'Ecole Normale Sup\'erieure, ENS, Universit\'e PSL, CNRS, Sorbonne Universit\'e, 75005 Paris, France\\
$^{86}$ Observatoire de Paris, Universit\'e PSL, Sorbonne Universit\'e, LERMA, 750 Paris, France\\
$^{87}$ Astrophysics Group, Blackett Laboratory, Imperial College London, London SW7 2AZ, UK\\
$^{88}$ SISSA, International School for Advanced Studies, Via Bonomea 265, 34136 Trieste TS, Italy\\
$^{89}$ INFN, Sezione di Trieste, Via Valerio 2, 34127 Trieste TS, Italy\\
$^{90}$ Departamento de Astrof\'{i}sica, Universidad de La Laguna, 38206, La Laguna, Tenerife, Spain\\
$^{91}$ Dipartimento di Fisica e Scienze della Terra, Universit\'a degli Studi di Ferrara, Via Giuseppe Saragat 1, 44122 Ferrara, Italy\\
$^{92}$ Istituto Nazionale di Fisica Nucleare, Sezione di Ferrara, Via Giuseppe Saragat 1, 44122 Ferrara, Italy\\
$^{93}$ Institut de Physique Th\'eorique, CEA, CNRS, Universit\'e Paris-Saclay 91191 Gif-sur-Yvette Cedex, France\\
$^{94}$ NASA Ames Research Center, Moffett Field, CA 94035, USA\\
$^{95}$ INAF, Istituto di Radioastronomia, Via Piero Gobetti 101, 40129 Bologna, Italy\\
$^{96}$ Universit\'e C\^{o}te d'Azur, Observatoire de la C\^{o}te d'Azur, CNRS, Laboratoire Lagrange, Bd de l'Observatoire, CS 34229, 06304 Nice cedex 4, France\\
$^{97}$ Institute for Theoretical Particle Physics and Cosmology (TTK), RWTH Aachen University, 52056 Aachen, Germany\\
$^{98}$ Institute for Astronomy, University of Hawaii, 2680 Woodlawn Drive, Honolulu, HI 96822, USA\\
$^{99}$ Department of Physics \& Astronomy, University of California Irvine, Irvine CA 92697, USA\\
$^{100}$ University of Lyon, UCB Lyon 1, CNRS/IN2P3, IUF, IP2I Lyon, France\\
$^{101}$ INFN-Sezione di Genova, Via Dodecaneso 33, 16146, Genova, Italy\\
$^{102}$ Aix-Marseille Universit\'e, CNRS, CNES, LAM, Marseille, France\\
$^{103}$ Department of Astronomy \& Physics and Institute for Computational Astrophysics, Saint Mary's University, 923 Robie Street, Halifax, Nova Scotia, B3H 3C3, Canada\\
$^{104}$ University Observatory, Faculty of Physics, Ludwig-Maximilians-Universit{\"a}t, Scheinerstr. 1, 81679 Munich, Germany\\
$^{105}$ Ruhr University Bochum, Faculty of Physics and Astronomy, Astronomical Institute (AIRUB), German Centre for Cosmological Lensing (GCCL), 44780 Bochum, Germany\\
$^{106}$ Department of Physics and Astronomy, Vesilinnantie 5, 20014 University of Turku, Finland\\
$^{107}$ Univ. Grenoble Alpes, CNRS, Grenoble INP, LPSC-IN2P3, 53, Avenue des Martyrs, 38000, Grenoble, France\\
$^{108}$ Centre de Calcul de l'IN2P3, 21 avenue Pierre de Coubertin 69627 Villeurbanne Cedex, France\\
$^{109}$ Dipartimento di Fisica, Sapienza Universit\`a di Roma, Piazzale Aldo Moro 2, 00185 Roma, Italy\\
$^{110}$ Centro de Astrof\'{\i}sica da Universidade do Porto, Rua das Estrelas, 4150-762 Porto, Portugal\\
$^{111}$ Dipartimento di Fisica - Sezione di Astronomia, Universit\'a di Trieste, Via Tiepolo 11, 34131 Trieste, Italy\\
$^{112}$ Institute of Cosmology and Gravitation, University of Portsmouth, Portsmouth PO1 3FX, UK\\
$^{113}$ Department of Mathematics and Physics E. De Giorgi, University of Salento, Via per Arnesano, CP-I93, 73100, Lecce, Italy\\
$^{114}$ INFN, Sezione di Lecce, Via per Arnesano, CP-193, 73100, Lecce, Italy\\
$^{115}$ INAF-Sezione di Lecce, c/o Dipartimento Matematica e Fisica, Via per Arnesano, 73100, Lecce, Italy\\
$^{116}$ CEA Saclay, DFR/IRFU, Service d'Astrophysique, Bat. 709, 91191 Gif-sur-Yvette, France\\
$^{117}$ Institute for Computational Science, University of Zurich, Winterthurerstrasse 190, 8057 Zurich, Switzerland\\
$^{118}$ Higgs Centre for Theoretical Physics, School of Physics and Astronomy, The University of Edinburgh, Edinburgh EH9 3FD, UK\\
$^{119}$ Universit\'e St Joseph; Faculty of Sciences, Beirut, Lebanon\\
$^{120}$ Department of Astrophysical Sciences, Peyton Hall, Princeton University, Princeton, NJ 08544, USA}

\date{Received date / Accepted date }

\abstract{This work considers which higher-order effects in modelling the cosmic shear angular power spectra must be taken into account for \emph{Euclid}. We identify which terms are of concern, and quantify their individual and cumulative impact on cosmological parameter inference from \emph{Euclid}. We compute the values of these higher-order effects using analytic expressions, and calculate the impact on cosmological parameter estimation using the Fisher matrix formalism. We review 24 effects and find the following potentially need to be accounted for: the reduced shear approximation, magnification bias, source-lens clustering, source obscuration, local Universe effects, and the flat Universe assumption. Upon computing these explicitly, and calculating their cosmological parameter biases, using a maximum multipole of $\ell=5000$, we find that the magnification bias, source-lens clustering, source obscuration, and local Universe terms individually produce significant ($\,>0.25\sigma$) cosmological biases in one or more parameters, and accordingly must be accounted for. In total, over all effects, we find biases in $\Omega_{\rm m}$, $\Omega_{\rm b}$, $h$, and $\sigma_{8}$ of $0.73\sigma$, $0.28\sigma$, $0.25\sigma$, and $-0.79\sigma$, respectively, for flat $\Lambda$CDM. For the $w_0w_a$CDM case, we find biases in $\Omega_{\rm m}$, $\Omega_{\rm b}$, $h$, $n_{\rm s}$, $\sigma_{8}$, and $w_a$ of $1.49\sigma$, $0.35\sigma$, $-1.36\sigma$, $1.31\sigma$, $-0.84\sigma$, and $-0.35\sigma$, respectively; which are increased relative to the  $\Lambda$CDM due to additional degeneracies as a function of redshift and scale.}

  
  \authorrunning{A.C.~Deshpande et al.} 

\maketitle
\newpage
\section{\label{sec:level1}Introduction}
Our current best-in-class framework for parameterising the Universe, the Lambda cold dark matter ($\Lambda$CDM) model, leaves several open questions. A key component, yet to be fully explained, is the acceleration of the expansion of the Universe and its proposed driver: dark energy. A powerful tool for studying this is cosmic shear -- the distortion of the ellipticities we observe for distant galaxies, by weak gravitational lensing from the large-scale structure of the Universe \citep[LSS; see for example][]{DETFrep}.

To-date, the most recent generation of cosmic shear surveys \citep{HSCpowerspec, kids1000, DESY3res} has been able to achieve precision cosmology competitive with cosmic microwave background experiments, for a combination of $\sigma_8$ and $\Omega_{\rm m}$ \citep{Planck18}. Now, upcoming Stage IV surveys \citep{DETFrep} will probe a greater area and depth than previously possible. For example, telescopes such as \emph{Euclid}\footnote{\url{https://www.euclid-ec.org/}} \citep{EuclidRB}, \emph{Nancy Grace Roman}\footnote{\url{https://roman.gsfc.nasa.gov/}} \citep{WFIRSTpap}, and the Vera C. Rubin Observatory\footnote{\url{https://www.lsst.org/}} \citep{LSSTpap} will achieve more than an order-of-magnitude increase in precision over existing surveys \citep[][heareafter EC20]{ISTforecast}. We must therefore ensure that any sources of bias in our theoretical formalism are properly accounted for.

In this work, we consider the common approximations made, and effects typically neglected  when modelling the cosmic shear angular power spectrum. This is of importance when deriving cosmological parameters from a shear-only analysis, but also in a 3$\times$2pt analyses where modelling the weak lensing power spectrum sufficiently is also essential. Throughout the literature, these effects have been studied independently and using varying specifications. Here, we evaluate them in a consistent framework, and quantify their cumulative impact on cosmology inferred from \emph{Euclid}'s weak lensing probe. As a first step, we review the literature and pinpoint which terms are potentially significant, as well as those for which the impact on the shear power spectrum has not been evaluated. A comparison of the typical magnitudes of the studied corrections is given in Table \ref{tab:effectreview}. To create Table \ref{tab:effectreview} we manually read the published numbers from graphs in the referenced papers for the auto-correlation cosmic shear power spectrum for a redshift bin closest to $z=1$ for a in each paper. These values are \emph{only approximate} due to the inherent inaccuracy of reading value from graphs and varying assumptions the papers\footnote{Every paper made various slightly differing assumptions regarding survey area, depth and number density, in all cases details can be found in the references. In the case that only correction function analyses were available, not power spectrum, we performed a Hankel transform over the correlation functions of the graphs over the angular range available.}.   

From our survey of the literature we identify the following as potentially significant systematic effects requiring full analysis:
\begin{itemize}
\item Reduced shear approximation: the effect of assuming that the measured two-point statistics of reduced shear are equal to those of the shear field \citep{Shapiro09, KrauseHirataRS}. It has been previously shown that this approximation will require relaxation for \emph{Euclid} \citep{EucRSMB},
\item Magnification bias: the change in the observed number density of sources, due to galaxies at the flux limit of the survey having their flux increased or decreased due to magnification by lensing \citep{MBorig}. This effect has also shown to significantly bias cosmological information from \emph{Euclid} \citep{EucRSMB, MBSims} if not accounted for. Additionally, magnification bias must also be accounted for in probes of galaxy clustering. Its impact on the \emph{Euclid} galaxy clustering probe is discussed in \cite{EuclidMagGCWL}.
\item Source-lens clustering: the intrinsic clustering of source galaxies correlated with the density field \citep{BernSLC, HamanaSLC, YuSLC}. Typically, it is assumed source galaxies are distributed homogeneously across the sky.
\item Source obscuration: A reduction in the observed galaxy distribution due to closely-spaced and blended or overlapping source galaxies \citep{SourceObs}.
\item Local Universe effects: a possible bias in our measurements of summary statistics of the LSS due to residing in a region with a higher-than-average density \citep{LUOrig, LUHall}. 
\item Flat Universe assumption: the impact of assuming that non-flat geometries are sufficiently well represented by modifying the expression for comoving distance, and neglecting the additional change in the lensing kernels used to calculate the shear power spectrum \citep{Spatflatpap}.
\end{itemize}

We made this determination by first excluding the terms which are fourth-order in the lens potential or higher, as these have consistently been shown to be sub-dominant \citep{CoorayHuforder, ShapCooray}. Among these are time delay-lens coupling, and deflection-deflection coupling \citep{DopplerBernardeau}, which result from foregoing the small-angle and thin lens approximations and solving the Sachs equation explicitly. Similarly, fourth-order correction terms resulting from relaxing the Born approximation and accounting for line-of-sight coupling of two foreground lenses are negligible \citep{ShapCooray}. Additionally, fourth-order reduced shear corrections \citep{KrauseHirataRS} were also neglected. Re-enforcing the sub-dominance of these terms is the fact that the standard reduced shear correction matches forward models and N-body simulations sufficiently well \citep{DodelsonRS, EucRSMB}. We further excluded fourth-order and higher terms resulting from the contribution of dark energy pressure to the lensing potential \citep{lensingbylambda}.

Furthermore, we neglected finite beam corrections \citep{finitebeamcoe,2019PhRvD..99b3525F}, that manifest as a fractional correction to the lensing power spectra on very small scales which is approximately $-(1/3)(\ell\theta)^2$, where $\ell$ is the angular multipole of the power spectrum and $\theta$ the mean angular size of galaxies \citep[this was verified by][using ray-tracing simulations]{2021A&A...655A..54B}; for \emph{Euclid} this results in a fractional correction of $-(\ell/1.2\times 10^6)^2$ to the power spectrum. We also neglect the effects of spatially-varying survey depth \citep{heydenreichspatdepth}, which can be accounted for directly in the covariance matrix through forward-modelling \citep{vardepthfix}.

Of the remaining effects, we then neglected those significantly smaller than the reduced shear correction. This proves a good comparison because this correction has been consistently demonstrated to produce biases close to the significance threshold \citep{Shapiro09, EucRSMB}. Accordingly, we neglected the impact of the Doppler-shift of galaxies on their two-point statistics. Due to the inhomogeneity of the Universe, galaxies have peculiar velocities which affect the measurement of their redshifts. If this is taken into account, it results in an additional contribution to the reduced shear \citep{DopplerBernardeau}. As can be seen in Table \ref{tab:effectreview}, this effect can be safely neglected as it is two or three orders of magnitude below the reduced shear correction. Additionally, the cosmological parameter biases resulting from it are two orders of magnitude below the level of concern \citep{DopplerDeshpande}.

Likewise, the effect of unequal-time correlators was neglected, as the resulting correction to the angular power spectrum is more than four orders of magnitude smaller than that for the reduced shear \citep{Unequaltimecorr}; as illustrated in Table \ref{tab:effectreview}. This correction is a consequence of relaxing the equal-time approximation, which approximates the cross-correlation matter power spectrum evaluated at different times as either the power spectrum at a fixed time, or by a geometric mean.

We also did not need to explicitly evaluate the effects of relaxing the Limber, and flat-sky approximations, as the combined corrections for these are an order-of-magnitude smaller than the reduced shear correction over the majority of the range of scales observed by \emph{Euclid} \citep{limitsofshear17}. This is again demonstrated in Table \ref{tab:effectreview}. The Limber approximation considers only correlations in the plane of the sky as contributing to the lensing signal, and projects others onto the plane of the sky by replacing spherical Bessel functions with Delta functions \citep{Limberorig, KaiserLimber, ExtendedLimber}.

\begin{table*}[ht]
\centering
\caption{List of higher-order correction terms to the shear angular power spectrum resulting from relaxing approximations. To illustrate their typical sizes and facilitate comparison, the values of these correction terms at redshift $z\sim1$ are also stated here. These are provided for $\ell$-modes 10, 100, and 1000, and as a percentage of sample variance i.e. $100\delta C_\ell/$(sample variance). Sample variance here is calculated using the definition of \cite{Kaiser92}, see equation (\ref{eq:samplevar}). These values are taken from the available literature for the case of Stage IV cosmic shear experiments. Unavailable values are represented by `N/A'. Corrections which  have a functional form that is fourth-order in lensing potential, and therefore sub-dominant, are denoted by O($\phi^4$). The references provided refer to the values stated where available, or to the work describing the correction where explicit values are not available. For the case where the absolute value of an effect is more than four orders of magnitude smaller than the sample variance, it is denoted by < 0.01. The sign  denotes whether the effect is to decrease the power (a negative sign) or to increase (no sign). The rows are ordered in decreasing amplitude for $\ell=1000$. These numbers were read from graphs published in the referenced papers, due to the inherent inaccuracy of this approach we quote only one decimal place; in the case that only correlation functions were provide these were converted to power spectrum results using a Hankel transform over the quoted angular range.}
\label{tab:effectreview}
\begin{tabular}{c c c c c}
\hline\hline
Correction & Reference & \multicolumn{3}{c}{Percentage of Sample Variance}\\
& & $\ell=10$ & $\ell=100$ & $\ell=1000$ \\
\hline
Source-lens clustering & \cite{YuSLC} & N/A & 6.0 & 57.2\\
Reduced shear + magnification bias & \cite{EucRSMB} & 0.4 & 1.8 & 15.3\\
Post-Limber reduced shear & \cite{postlimbrs} & 0.2 & 0.6 & 1.9\\
Non-linear ellipticity-shear relation & \cite{KrauseHirataRS} & 0.2 & 0.6 & 1.9\\
Limber + flat-sky & \cite{limitsofshear17} & 9.8 & 3.0 & 1.0 \\
Local Universe effects & \cite{LUHall} & 7.8 & 24.2 & N/A\\
Higher-order reduced shear & \cite{KrauseHirataRS} & O($\phi^4$) & O($\phi^4$) & O($\phi^4$)\\
Time delay-lens coupling & \cite{DopplerBernardeau} & O($\phi^4$) & O($\phi^4$) & O($\phi^4$) \\
Deflection-deflection coupling & \cite{DopplerBernardeau} & O($\phi^4$) & O($\phi^4$) & O($\phi^4$) \\
Born approximation & \cite{CoorayHuforder} & O($\phi^4$) & O($\phi^4$) & O($\phi^4$) \\
Lensing by dark energy pressure & \cite{lensingbylambda} & O($\phi^4$) & O($\phi^4$) & O($\phi^4$) \\
Second-order speed-of-light & \cite{CuestaLazaroManyCorrs} &  < 0.01 & < 0.01 & < 0.01 \\
Temporal-Born approximation & \cite{CuestaLazaroManyCorrs} &  < 0.01 &  < 0.01 &  < 0.01 \\
Finite-beam corrections & \cite{2019PhRvD..99b3525F} &  < 0.01 &  < 0.01 &  < 0.01 \\
Doppler-shift & \cite{DopplerDeshpande} &  < 0.01 &  < 0.01 &  < 0.01 \\
Unequal-time correlators & \cite{Unequaltimecorr} &  < 0.01 &  < 0.01 &  < 0.01 \\
Sachs-Wolfe effect & \cite{CuestaLazaroManyCorrs}  &  < 0.01 &  < 0.01 &  < 0.01 \\
Integrated Sachs-Wolfe effect & \cite{CuestaLazaroManyCorrs}  &  < 0.01 &  < 0.01 &  < 0.01 \\
Flexion correction & \cite{flexionpap}  & N/A & N/A & N/A \\
Flat-geometry assumption & \cite{Spatflatpap} & $-2.0$ & $-6.0$ & $-19.1$ \\
Source obscuration & \cite{SourceObs} & $-$2.0 & $-6.0$ & $-19.1$\\
Spatially-varying survey depth & \cite{heydenreichspatdepth} & $-5.9$ & $-18.1$ & $-57.2$ \\
\hline\hline
\end{tabular}
\end{table*}

Additionally, the Limber approximation is also employed when computing higher-order corrections to the angular power spectrum, such as the reduced shear correction. In this work, we also did not relax the use of the Limber approximation here, as the cosmological parameter biases from this are safely negligible \citep{postlimbrs}.

Another series of corrections that we deemed safely negligible stem from corrections to the theoretical expressions describing light propagation \citep{CuestaLazaroManyCorrs}. Among these is the effect of second-order corrections to the effective speed of light. This relaxes the assumption that, as lensing potentials are small, the lensing effect can be studied in an effective Minkowskian spacetime and, accordingly, the effective speed of light need only be computed to the first-order. Similarly, a second correction to the effective speed of light presents itself from the energy-momentum tensor. Typically, this quantity is calculated under the assumption that lenses are moving slowly, so that the kinetic contribution to gravity can be ignored. Addressing this creates another correction to the angular power spectrum. 

A further effect is that the observed ellipticity is non-linearly related to the shear, but that a linear approximation is often made. For a discussion of the impact of the non-linear relation on the observed shear distribution see \cite{2014MNRAS.439.1909V}. The impact of this effect on the power spectrum (if one assumes a linear relation instead of the non-linear relation) is investigated in \cite{KrauseHirataRS} (Section 3.3) who find that the impact is three orders of magnitude smaller than the shear power spectrum\footnote{In fact there are two definitions of ellipticity: third eccentricity and third flattening \citep[see e.g.][]{2014MNRAS.439.1909V} that relate the observed ellipticity to the shear in different ways. \cite{KrauseHirataRS} find that for third flattening the correction to the power spectrum is zero \cite[since the moments of the third flattening are exactly the moments of the reduced shear as shown by][]{1997A&A...318..687S}, but that for third eccentricity the effect is non-zero.}.

The temporal-Born approximation is another correction to the description of light propagation. While the correction for the standard Born approximation accounts for the spatial discrepancy between the true perturbed path of a photon from source to observer compared to the mathematically convenient straight one, this discrepancy also produces a temporal one. The photon on the perturbed path will at times be ahead of the photon on the idealised path, and at other times lag behind. Accordingly, the two would encounter different evolutionary stages of the LSS at different times, necessitating a correction in the two-point statistics. The remaining two corrections that fall under this umbrella are the corrections of the Sachs-Wolfe and integrated Sachs-Wolfe effects. The former describes the redshift of an emitted photon due to the source galaxy's gravitational potential, while the latter encodes the effect on the photon of interaction with the evolving gravitational potential along its path. All of these light propagation corrections are many orders of magnitude below the reduced shear correction, as can be seen in a representative example in Table \ref{tab:effectreview}.

Finally, we deemed the flexion corrections to be negligible without requiring explicit calculation. This additional correction term arises from the fact that, for larger sources, the image distortion consists of both shear and a higher-order component labelled flexion \citep{flexionpap}. This term should be negligible because its effect on the cosmic shear signal will be dependent on third-order or higher-order brightness moments.

Here, we did not consider biases arising from general modelling of unknown shape measurement systematic effects (i.e. multiplicative and additive biases). Instead, we focused only on these well defined theoretical assumptions. For more details on shape measurement effects, see \cite{ShapeMeas19, ShapeMeas20, ShapeMeas21, ShapeMeas22}. Additionally, we did not evaluate the additional selection effects of flux cuts and size cuts, as the former of these can be calibrated from deep fields, and size cuts are primarily a concern for ground-based telescopes, rather than space-based ones.

In this work, we also did not examine the impact of neglecting effects that are already well-established as requiring evaluation, i.e. photometric redshift uncertainties, intrinsic alignments (IA) modelling, baryonic feedback, and modelling of the non-linear component of the matter power spectrum. Determining the exact specification for these is outside of the scope of this work, and each of those effects requires its own through investigation.

This work is structured as follows: In Sect. \ref{sec:2}, we detail the theoretical formalism used. We review the basic, first-order cosmic shear angular power spectrum calculation. Then, the expressions for the six correction terms of interest are detailed. We also describe the Fisher matrix formalism used to predict cosmological parameter constraints and biases. In Sect. \ref{sec:3}, we discuss the modelling and computational specifics used in this work. Finally, we discuss our results in Sect. \ref{sec:4}. We show the cosmological parameter biases that result from neglecting the studied corrections, and discuss their implications for \emph{Euclid}.

\section{\label{sec:2}Theoretical formalism}

Here, we begin by reviewing the standard first-order calculation of the cosmic shear angular power spectrum. Additional contributions to the lensing signal resulting from IAs and shot-noise are then described. We then detail the analytical forms of the six corrections requiring full evaluation: reduced shear, magnification bias, source-lens clustering, source obscuration,  local Universe effects, and the flat Universe assumption. Finally, we review the Fisher matrix formalism used to predict cosmological parameter constraints and biases.

\subsection{\label{subsec:angpow}The first-order cosmic shear calculation}

As a consequence of weak gravitational lensing by the LSS, the observed ellipticity of distant galaxies is distorted. This change is dependent on the reduced shear, $g$, according to
\begin{equation}
    \label{eq:redshear}
    g^\alpha(\boldsymbol{\theta})= \frac{\gamma^\alpha(\boldsymbol{\theta})}{1-\kappa(\boldsymbol{\theta})}\, ,
\end{equation}
where $\boldsymbol{\theta}$ is the position of the galaxy on the sky, $\gamma$ is the spin-2 shear with index $\alpha \in \{1,2\}$ which describes the anisotropic stretching that turns circular distributions of light elliptical, and $\kappa$ is the convergence -- responsible for the isotropic change in the size of the image. Since in the weak lensing regime $|\kappa| \ll 1$, it is standard practice to make the reduced shear approximation, whereby
\begin{equation}
    \label{eq:RSA}
    g^\alpha(\boldsymbol{\theta}) \approx \gamma^\alpha(\boldsymbol{\theta})\, .
\end{equation}

Additionally, the convergence is a projection of the density contrast of the Universe, $\delta$, along the line-of-sight over comoving distance, $\chi$, to the comoving distance to the horizon, $\chi_{\rm h}$. For a particular tomographic redshift bin $i$, it is mathematically described by
\begin{equation}
    \label{eq:convergence}
    \kappa_i(\boldsymbol{\theta})=\int_{0}^{\chi_{\rm h}} {\rm d}\chi\:\delta[S_K(\chi)\boldsymbol{\theta},\, \chi]\:W_i(\chi)\, ,
\end{equation}
where $S_K$ is a function that encodes the effect of the curvature of the Universe, $K$, on comoving distances according to
\begin{equation}
    \label{eq:SK}
    S_K(\chi) = \begin{cases}
    |K|^{-1/2}\sin(|K|^{-1/2}\chi) & \text{\small{$K>0$ (closed Universe)}}\\
    \chi & \text{\small{$K=0$ (flat Universe)}}\\
    |K|^{-1/2}\sinh(|K|^{-1/2}\chi) & \text{\small{$K<0$ (open Universe)}}\,.
  \end{cases}
\end{equation}
We remind the reader that for the quantity $\delta[S_K(\chi)\boldsymbol{\theta},\chi]$ in equation (\ref{eq:convergence}) the second $\chi$ means not only that there is an evaluation at a comoving radius $\chi$, but also at a conformal time $\eta = \eta_0 - \chi$, meaning that all the integration over $\chi$ in this the paper are performed down the background light cone.

The $W_i(\chi)$ in Eq.~(\ref{eq:convergence}) is the lensing projection kernel for tomographic bin $i$. It takes the form
\begin{align}
    \label{eq:Wi}
    W_i(\chi) &= \frac{3}{2}\Omega_{\rm m}\frac{H_0^2}{c^2}\frac{S_K(\chi)}{a(\chi)}\int_{\chi}^{\chi_{\rm h}}{\rm d}\chi'\:n_i(\chi')\frac{S_K(\chi'-\chi)}{S_K(\chi')}\, ,
\end{align}
which is dependent on the dimensionless present-day matter density of the Universe $\Omega_{\rm m}$, the speed of light in a vacuum $c$, the Hubble constant $H_0$, the scale factor of the Universe $a(\chi)$, and the probability distribution of galaxies within redshift bin $i$ $n_i(\chi)$.

The spin-2 shear is directly related to the convergence in spherical-harmonic space. For a specified lensing mass distribution, assuming the flat-sky and prefactor-unity approximations \citep{limitsofshear17}, and under the small-angle limit, this relationship takes the form
\begin{equation}
    \label{eq:fourier}
    \widetilde{\gamma}_i^\alpha(\boldsymbol{\ell})= T^\alpha(\boldsymbol{\ell})\,\widetilde{\kappa}_i(\boldsymbol{\ell})\, ,
\end{equation}
where $\boldsymbol{\ell}$ is the spherical-harmonic conjugate of $\boldsymbol{\theta}$, with magnitude $\ell$ and angular component $\phi_\ell$. The functions $T^\alpha$ are two trigonometric weighting functions corresponding to each of the shear components. These take the form:
\begin{align}
    \label{eq:Trigfunc1}
    T^1(\boldsymbol{\ell}) &= \cos(2\phi_\ell)\, ,\\
    \label{eq:Trigfunc2}
    T^2(\boldsymbol{\ell}) &= \sin(2\phi_\ell)\, .
\end{align}

In the case of an arbitrary shear field, for example a field constructed from data, two linear combinations of the individual shear components are pertinent. Specifically, these are a divergence-free $B$-mode, and a curl-free $E$-mode:
\begin{align}
    \label{eq:Emode}
    \widetilde{E}_i(\boldsymbol{\ell})&=\sum_\alpha T^\alpha(\boldsymbol{\ell})\:\widetilde{\gamma}_i^\alpha(\boldsymbol{\ell})\, ,\\
    \label{eq:Bmode}
    \widetilde{B}_i(\boldsymbol{\ell})&=\sum_\alpha \sum_\beta \varepsilon^{\alpha\beta}\,T^\alpha(\boldsymbol{\ell})\:\widetilde{\gamma}_i^\beta(\boldsymbol{\ell})\, .
\end{align}
Here, the summations are over the shear components, and $\varepsilon^{\alpha\beta}$ is the Levi-Civita symbol in the two-dimensional case; such that: $\varepsilon^{11}=\varepsilon^{22}=0$ and $\varepsilon^{12}=-\varepsilon^{21}=1$.

Assuming that higher-order systematic effects in the data have been accounted for, the $B$-mode of Eq.~(\ref{eq:Bmode}) vanishes. For the remaining $E$-mode, observables of interest are defined in the form of angular auto and cross-correlation power spectra, $C_{\ell;ij}^{\gamma\gamma}$, such that
\begin{equation}
    \label{eq:powerspecdef}
    \left<\widetilde{E}_i(\boldsymbol{\ell})\widetilde{E}_j(\boldsymbol{\ell}')\right> = (2\pi)^2\,\delta_{\rm D}^{(2)}(\boldsymbol{\ell}+\boldsymbol{\ell}')\,C_{\ell;ij}^{\gamma\gamma}\, ,
\end{equation}
where the angular brackets on the left-hand-side denote the ensemble average, which under the assumption of ergodicity becomes a spatial average, and $\delta_{\rm D}^{(2)}$ is the Dirac delta for two dimensions. Under the extended Limber approximation \citep{ExtendedLimber}, where $k=(\ell + 1/2)/S_K(\chi)$, the power spectra themselves are further defined as
\begin{equation}
    \label{eq:Cl}
    C_{\ell;ij}^{\gamma\gamma} = \int_0^{\chi_{\rm h}}{\rm d}\chi\frac{W_i(\chi)W_j(\chi)}{S^{\,2}_K(\chi)}P_{\delta\delta}(k, \chi)\, ,
\end{equation}
where $P_{\delta\delta}$ is the three-dimensional matter power spectrum, and $k$ is the magnitude of the spatial momentum vector $\boldsymbol{k}$ which also shares the angular component $\phi_\ell$. Detailed reviews of this standard calculation can be found in \cite{Kilbinger15, Munshirev, BartSchneiRev}.

\subsection{\label{subsec:IAshotnoise}Intrinsic alignments and shot noise}

When the angular power spectra are actually measured from surveys of galaxies, they contain non-lensing signals together with the pertinent cosmic shear power spectra. It is necessary to model each of these components to ensure accurate cosmological inference. A key non-lensing contribution arises from the fact that galaxies forming close to each other are forming in a similar tidal environment. Consequently, they have intrinsically correlated alignments \citep{IA1, IA2, IA3}.

The observed ellipticity of an individual source, $\epsilon$, can then, to first-order, be written as a combination of its underlying ellipticity in the absence of any cosmic shear or IA, $\epsilon^{\rm s}$, the cosmic shear, $\gamma=\gamma^1+{\rm i}\gamma^2$, and the effect of IA, $\epsilon^{\rm I}$ according to
\begin{equation}
    \label{eq:galelip}
    \epsilon = \epsilon^{\rm s} + \gamma + \epsilon^{\rm I}\, .
\end{equation}
The angular power spectra corresponding to this observed ellipticity, $C_{\ell;ij}^{\epsilon\epsilon}$, are then the sum of contributions resulting from its components,
\begin{equation}
    \label{eq:ObsCl}
    C_{\ell;ij}^{\epsilon\epsilon} = C_{\ell;ij}^{\gamma\gamma} + C_{\ell;ij}^{\gamma{\rm I}} + C_{\ell;ij}^{{\rm I}\gamma} + C_{\ell;ij}^{\rm II} + N_{\ell;ij}^\epsilon\, ,
\end{equation}
in which $C_{\ell;ij}^{\gamma\gamma}$ are the cosmic shear angular power spectra defined in Eq.~(\ref{eq:Cl}); in all cases the notation denotes $z_i \leq z_j$. The $C_{\ell;ij}^{\gamma{\rm I}}$ are the angular power spectra of correlations between foreground shear and background IA, which are only non-zero if photometric redshift estimates result in the scattering of observed redshifts between bins. On the other hand, the $C_{\ell;ij}^{{\rm I}\gamma}$ arise from the correlation between background shear and foreground IA, and the $C_{\ell;ij}^{\rm II}$ represent the auto-correlation of the IA; both must be accounted for. To accomplish this, the non-linear alignment (NLA) model \citep{IA_NLA} can be employed. Under this model, these IA spectra take the form
\begin{align}
    \label{eq:cllig}
    C_{\ell;ij}^{{\rm I}\gamma} &= \int_0^{\chi_{\rm h}}\frac{{\rm d}\chi}{S^{\,2}_K(\chi)}[W_i(\chi)n_j(\chi)+n_i(\chi)W_j(\chi)]P_{\delta {\rm I}}(k, \chi)\, ,\\
    \label{eq:clli}
    C_{\ell;ij}^{\rm II} &= \int_0^{\chi_{\rm h}}\frac{{\rm d}\chi}{S^{\,2}_K(\chi)}n_i(\chi)n_j(\chi)\,P_{\rm II}(k, \chi)\, ,
\end{align}
which, in a similar manner to the shear power spectra, are projections of three-dimensional IA power spectra, $P_{\delta {\rm I}}$ and $P_{\rm II}$. Both of these are related to the matter power spectrum as follows:
\begin{align}
    \label{eq:pdi}
    P_{\delta {\rm I}}(k, \chi) &= \bigg[-\frac{\mathcal{A}_{\rm IA}\mathcal{C}_{\rm IA}\Omega_{\rm m}}{D(\chi)}\bigg]\:\:P_{\delta\delta}(k,\chi)\, ,\\
    \label{eq:pii}
    P_{\rm II}(k, \chi) &= \bigg[-\frac{\mathcal{A}_{\rm IA}\mathcal{C}_{\rm IA}\Omega_{\rm m}}{D(\chi)}\bigg]^2P_{\delta\delta}(k,\chi)\, ,
\end{align}
where the product of $\mathcal{A}_{\rm IA}$ and $\mathcal{C}_{\rm IA}$ is a free parameter typically set by fitting to simulations or data, and $D(\chi)$ is the density perturbation growth factor. We note that the NLA model is a limited description of IAs, and accordingly has its own associated modelling uncertainties. Extensions of this model have been proposed \citep[see e.g. ][EC20]{FortunaIA}. However, investigating the modelling of IAs for \emph{Euclid} in detail is out of the scope of this work, and necessitates a separate future investigation of itself. 

Of the terms in Eq.~(\ref{eq:ObsCl}), $N_{\ell;ij}^\epsilon$ remains; this shot noise term arises from the zero-lag autocorrelation of the unlensed, uncorrelated source ellipticity $\epsilon^{\rm s}$ in Eq.~(\ref{eq:galelip}) \citep[see e.g.][equation 4]{1999ApJ...522L..21H}. For a survey with equi-populated tomographic redshift bins, such as \emph{Euclid} (EC20), this is expressed by
\begin{align}
    \label{eq:shotnoise}
    N_{\ell;ij}^\epsilon = \frac{\sigma_\epsilon^2}{\bar{n}_{\rm g}/N_{\rm bin}}\delta_{ij}^{\rm K}\, ,
\end{align}
within which $\sigma_\epsilon^2$ is the variance of the observed ellipticities in the survey, $\bar{n}_{\rm g}$ is the surface density of galaxies in the survey, $N_{\rm bin}$ is the survey's number of tomographic redshift bins, and $\delta_{ij}^{\rm K}$ is the Kronecker delta -- which here indicates that the shot noise vanishes for cross-correlation spectra, as the ellipticities of galaxies at differing redshifts should not be correlated.

\subsection{\label{subsec:reducedshear}The reduced shear approximation}

While relaxing the reduced shear approximation completely and explicitly is intractable, this can be sufficiently well modelled by applying a second-order Taylor expansion \citep{DodelsonRS, Shapiro09, KrauseHirataRS, EucRSMB} to Eq.~(\ref{eq:redshear}), resulting in
\begin{equation}
    \label{eq:gexpan}
    g^\alpha(\boldsymbol{\theta})=\gamma^\alpha(\boldsymbol{\theta})+(\gamma^\alpha\kappa)(\boldsymbol{\theta})+\mathcal{O}(\kappa^3)\, .
\end{equation}

Computing the angular $E$-mode power spectra using this expanded expression results in the standard two-point expression of Eq.~(\ref{eq:powerspecdef}), plus three-point terms. These additional terms, $\delta\braket{\widetilde{E}_i(\boldsymbol{\ell})\widetilde{E}_j(\boldsymbol{\ell}')}$, are given by
\begin{align}
    \label{eq:ecorr}
    \delta\braket{\widetilde{E}_i(\boldsymbol{\ell})\widetilde{E}_j(\boldsymbol{\ell}')} &=  \sum_\alpha \sum_\beta T^\alpha(\boldsymbol{\ell})T^\beta(\boldsymbol{\ell}')\braket{\widetilde{(\gamma^\alpha\kappa)}_i(\boldsymbol{\ell})\:\widetilde{\gamma}_j^\beta(\boldsymbol{\ell'})} \nonumber\\ &+ T^\alpha(\boldsymbol{\ell}')T^\beta(\boldsymbol{\ell})\braket{\widetilde{(\gamma^\alpha\kappa)}_j(\boldsymbol{\ell}')\:\widetilde{\gamma}_i^\beta(\boldsymbol{\ell})} \nonumber \\
    &= (2\pi)^2\,\delta_{\rm D}^{(2)}(\boldsymbol{\ell}+\boldsymbol{\ell}')\:\delta C^{\rm RS}_{\ell;ij}\, ,
\end{align}
where $\delta C^{\rm RS}_{\ell;ij}$ is the corresponding correction to $C_{\ell;ij}^{\gamma\gamma}$, and is given by
\begin{align}
    \label{eq:dCl_RS}
    \delta C^{\rm RS}_{\ell;ij} &= \int_0^\infty\frac{{\rm d}^2\boldsymbol{\ell}'}{(2\pi)^2}\cos(2\phi_{\ell'})B_{ij}^{\kappa\kappa\kappa}(\boldsymbol{\ell}, \boldsymbol{\ell}', -\boldsymbol{\ell}-\boldsymbol{\ell}')\, ,
\end{align}
where we are always free to choose a coordinate system such that $\phi_{\ell} = 0$, and accordingly the correction only depends on the magnitude, $\ell$. It depends on the two-redshift convergence bispectrum, $B_{ij}^{\kappa\kappa\kappa}$, which is the three-point counterpart of the convergence power spectrum. Higher-order terms in the Taylor expansion of Eq.~(\ref{eq:gexpan}) would here result in corrections dependent on the matter trispectrum, as well as Wick contraction terms of O($P^2_{\delta\delta}$). Both types of terms have been shown to be sub-dominant \citep{CoorayHuforder, ShapCooray, DodelsonRS, KrauseHirataRS, EucRSMB}. The latter type of term, although of the same perturbative order in the power spectrum as the bispectrum, is still of O($W(\chi)^4$), and given that typically $\chi W(\chi) \ll 1$, it will still be significantly smaller than the correction of Eq.~(\ref{eq:dCl_RS}).

Additionally, just as the convergence power spectrum is the projection of the matter power spectrum, the convergence bispectrum is analogously the projection of the matter bispectrum, $B_{\delta\delta\delta}$. Under the Limber approximation, this takes the form
\begin{align}
    \label{eq:bispecK}
    B_{ij}^{\kappa\kappa\kappa}(\boldsymbol{\ell}_1, \boldsymbol{\ell}_2, \boldsymbol{\ell}_3) &= B_{iij}^{\kappa\kappa\kappa}(\boldsymbol{\ell}_1, \boldsymbol{\ell}_2, \boldsymbol{\ell}_3) + B_{ijj}^{\kappa\kappa\kappa}(\boldsymbol{\ell}_1, \boldsymbol{\ell}_2, \boldsymbol{\ell}_3)\nonumber\\ &=\int_0^{\chi_{\rm h}}\frac{{\rm d}\chi}{S^{\,4}_K(\chi)}W_i(\chi)W_j(\chi)[W_i(\chi)+W_j(\chi)] \nonumber\\
    &\times B_{\delta\delta\delta}(\boldsymbol{k}_1,\boldsymbol{k}_2,\boldsymbol{k}_3,\chi)\, .
\end{align}
For a relaxation of the Limber approximation see \cite{postlimbrs}.

It should also be noted that the use of the reduced shear approximation can produce a $B$-mode signal contribution. However, it has been demonstrated that this is negligible \citep{Bmodespap}.

\subsection{\label{subsec:slc}Source-lens clustering}

Since, in practice, cosmic shear is only measured where galaxies are present, care must be taken to account for biases from any correlations between background source galaxies and the foreground lensing field. Given that, in reality, tomographic bins must be wide enough to include a sufficient number of galaxies so that shape-measurement noise is minimised, there will be overlap between the source and lensing distributions. The situation is further aggravated by broadening of bins due to photometric redshift uncertainties.

As a consequence of this effect, the observed number density of galaxies used in a given estimator which determines the shear angular power spectra from data is correlated with the intrinsic source galaxy overdensity, $\delta^g_i$, such that \citep{BernSLC, HamanaSLC, RSMBcombpap}
\begin{align}
    \label{eq:n_slc}
    n_i^{\rm obs}(\boldsymbol{\theta}, \chi) = n_i(\chi)\,[1 + \delta^g_i(\boldsymbol{\theta})]\, .
\end{align}
Accordingly, the shear used in the theoretical formalism for inference, is similarly replaced with an `observed' shear,
\begin{align}
    \label{eq:shear_slc}
    \gamma^\alpha_{{\rm obs}; i}(\boldsymbol{\theta}) &= \gamma^\alpha_i(\boldsymbol{\theta}) +  \gamma^\alpha_i(\boldsymbol{\theta})\,\delta^g_i(\boldsymbol{\theta})\, .
\end{align}
This is similar in form to the Taylor expansion of the reduced shear expressed in Eq.~(\ref{eq:gexpan}), and results in an analogous correction term, $\delta C_{\ell; ij}^{\rm SLC}$, to the angular power spectra,
\begin{align}
    \label{eq:dCl_slc}
    \delta C^{\rm SLC}_{\ell;ij} &= \int_0^\infty\frac{{\rm d}^2\boldsymbol{\ell}'}{(2\pi)^2}\cos(2\phi_{\ell'})B_{ij}^{\kappa\delta^g \kappa}(\boldsymbol{\ell}, \boldsymbol{\ell}', -\boldsymbol{\ell}-\boldsymbol{\ell}')\, ,
\end{align}
where $B_{ij}^{\kappa\delta^g \kappa}$ is now the two-redshift convergence-galaxy bispectrum. By adopting a linear galaxy bias model (so that $\delta_g = b\,\delta$) as used in EC20, and noting that $\delta^g$ is the 2D projection of $\delta_g$, the convergence-galaxy bispectrum can also be expressed as a projection of the matter bispectrum
\begin{align}
    \label{eq:bispec_slc}
    B_{ij}^{\kappa\delta^g \kappa}(\boldsymbol{\ell}_1, \boldsymbol{\ell}_2, \boldsymbol{\ell}_3) &= B_{iij}^{\kappa\delta^g \kappa}(\boldsymbol{\ell}_1, \boldsymbol{\ell}_2, \boldsymbol{\ell}_3) + B_{ijj}^{\kappa\delta^g \kappa}(\boldsymbol{\ell_1}, \boldsymbol{\ell}_2, \boldsymbol{\ell}_3)\nonumber\\ &=\int_0^{\chi_{\rm h}}\frac{{\rm d}\chi}{S^{\,4}_K(\chi)}
    [b_i\,n_i(\chi)+b_j\,n_j(\chi)]W_i(\chi)W_j(\chi) \nonumber\\
    &\times B_{\delta\delta\delta}(\boldsymbol{k}_1,\boldsymbol{k}_2,\boldsymbol{k}_3,\chi)\, ,
\end{align}
where $b_i$ and $b_j$ are the galaxy biases for tomographic bins $i$ and $j$, respectively. While more complex models of the galaxy bias exist, we proceed with the linear bias in this work, in order to mitigate the already significant computational load of these three-point terms. We note that modelling the galaxy bias requires more complexity at smaller scales, where the SLC effect is most relevant. When ultimately computing this term in the \emph{Euclid} cosmological analysis, the final \emph{Euclid} galaxy bias model should be used.
The linear galaxy bias for each tomographic bin is given by
\begin{align}
    \label{eq:lingalbias}
    b_{i} = \sqrt{1+\bar{z_i}}\, ,
\end{align}
where $\bar{z_i}$ is tomographic bin $i$'s central redshift. For a review of galaxy bias models, see \cite{galbiasrev}.

In addition to this contribution to the $E$-mode angular power spectra, source-lens clustering produces a $B$-mode signal as well. This term is comparable to the $E$-mode correction in magnitude, and accordingly, its detection in the absence of other $B$-mode contributions could allow for direct correction of the $E$-mode signal, rather than requiring the computation of Eq.~(\ref{eq:dCl_slc}). However, typical $B$-mode signals are dominated by other contributions \citep{Bmodespap,YuSLC}.

\subsection{\label{subsec:magbias}Magnification bias}

An additional consequence of gravitational lensing is that the density of galaxies observed by a particular survey is no longer representative of the true underlying galaxy density \citep{MBorig}. In particular, magnification resulting from the convergence modifies the density in two contrasting ways.

One manifestation of the effect is that individual sources are magnified, and as a consequence of this, their flux increases. Accordingly, any sources lying just beyond the flux limit of the survey may have their fluxes increased to the point of then being within the flux limit; increasing the observed density. As sources are magnified, the patch of sky around them too is magnified. This causes the second, competing manifestation. Within the magnified patch of sky, the galaxy density is reduced. The total effect, known as magnification bias, is dependent on the slope of the unlensed galaxy luminosity function. This assumes that the magnification $\mu > 1$.

Assuming that, on our scales of interest, fluctuations in the intrinsic galaxy overdensity are small, and taking into account that, for weak lensing, $|\kappa| \ll 1$, the observed galaxy overdensity for a given tomographic bin, $\delta^g_{{\rm obs}; i}$, is given by \citep{MBcorssource, RSMBcombpap}
\begin{equation}
    \label{eq:galover}
    \delta^g_{{\rm obs}; i}(\boldsymbol{\theta}) = \delta^g_i(\boldsymbol{\theta}) + (5s_i-2)\kappa_i(\boldsymbol{\theta})\, ,
\end{equation}
where $\delta^g_i$ is the intrinsic galaxy overdensity in the absence of magnification or any other systematic effects, and $s_i$ is the slope of the luminosity function for redshift bin $i$. This is given by the derivative of the cumulative galaxy number counts with respect to magnitude, $m$, evaluated at the survey's limiting magnitude, $m_{\rm lim}$ such that
\begin{equation}
    \label{eq:slope_lum}
    s_i = \frac{\partial{\rm log}_{10}\,\mathfrak{n}(\bar{z_i}, m)}{\partial m}\bigg|_{m_{\rm lim}}\, ,
\end{equation}
in which $\mathfrak{n}(\bar{z_i}, m)$ is the true, underlying distribution of galaxies, evaluated at the tomographic bin's central redshift, $\bar{z_i}$. Here, we have suppressed an additional dependence on the wavelength band in which the galaxy is observed. This should be considered when determining the slope from observational data.

Accordingly, Eq.~(\ref{eq:shear_slc}) gains an extra term
\begin{align}
    \label{eq:MBshear}
    \gamma^\alpha_{{\rm obs}; i} &= \gamma^\alpha_i(\boldsymbol{\theta}) +\gamma^\alpha_i(\boldsymbol{\theta})\,\delta^g_i(\boldsymbol{\theta}) + (5s_i-2)\,\gamma^\alpha_i(\boldsymbol{\theta})\kappa_i(\boldsymbol{\theta})\, .
\end{align}
This additional term is near-identical to the second term in Eq.~(\ref{eq:gexpan}), but for the prefactor of $(5s_i-2)$. Accordingly, it too spawns a correction to the angular power spectra. This correction for magnification bias, $\delta C_{\ell; ij}^{\rm MB}$, takes a similar form to the reduced shear correction of Eq.~(\ref{eq:dCl_RS}),
\begin{align}
    \label{eq:dCl_MB}
    \delta C^{\rm MB}_{\ell;ij} &= \int_0^\infty\frac{{\rm d}^2\boldsymbol{\ell}'}{(2\pi)^2}\cos(2\phi_{\ell'})[(5s_i-2)B_{iij}^{\kappa\kappa\kappa}(\boldsymbol{\ell}, \boldsymbol{\ell}', -\boldsymbol{\ell}-\boldsymbol{\ell}')\nonumber\\
    &+ (5s_j-2)B_{ijj}^{\kappa\kappa\kappa}(\boldsymbol{\ell}, \boldsymbol{\ell}', -\boldsymbol{\ell}-\boldsymbol{\ell}')]\, .
\end{align}

Given the similarity of the magnification bias correction to the source-lens clustering and reduced shear corrections, it too would produce a contribution to the $B$-mode signal. While this has not been explicitly evaluated, we would expect this term, as with its reduced shear and source-lens clustering counterparts, to be sub-dominant.

\subsection{\label{subsec:sourceobs}Source obscuration}

There is another systematic effect which can change the observed galaxy number density. Blending of close galaxy pairs can lead to multiple galaxies being discarded, or counted as a lower number than they are \citep{SourceObs}. The resulting change in the observed number density of galaxies, $\Delta\, n^{\rm SO}(z, \boldsymbol{\theta})$, can be modelled by
\begin{align}
    \label{eq:so_ndelt_full}
    \frac{\Delta\, n^{\rm SO}(z, \boldsymbol{\theta})}{n(z, \boldsymbol{\theta})}&=-\pi\left[(2\vartheta)^{2}\, n_{\text {tot }}+\frac{A\, n(z)}{2-\zeta}(2\vartheta)^{2-\zeta}\right]\,,
\end{align}
where $n_{\rm tot}$ is the total number density of galaxies at all redshifts, $n(z)$ is the observed density of galaxies at redshift $z$ ignoring source-lens clustering, we assume a redshift-independent radius $\vartheta$ for all galaxies as in \cite{SourceObs}, and $A$ and $\zeta$ are the amplitude and power-law index of a power-law model for the two-point galaxy angular correlation function. This expression is obtained by considering the probability that the centroid of another source lies within 2$\vartheta$ of a given one by integrating over the probability that another source centroid lies in an annulus of d$\theta$ around the centroid of another one.

Instead, we assume that the blending strategy for \emph{Euclid} will  account for blended pairs sufficiently well, such that the only obscuration of concern is substantial overlap; when the centroid of a source is behind another source (i.e. within $\vartheta$ rather than 2$\vartheta$). In this case, we are only concerned with the probability of this overlap. We note that in reality this blending has a complex interaction with shape measurement, but evaluating this is out of the scope of this work. Then, assuming that sources are approximately circular, the probability, ${\rm d}p$, of a galaxy at redshift $z$ overlapping with one at redshift $z'$ is
\begin{align}
    \label{eq:prob_so}
    {\rm d}p (z, z', \boldsymbol{\theta}) =  \pi\, \vartheta^2\,n(z', \boldsymbol{\theta})\, {\rm d}z \, .
\end{align}
Here it is also assumed that the expected number of galaxies overlapping with a given galaxy is $\ll 1$, such that the probability of at least one galaxy overlapping a given source (resulting in the removal of that source from the sample) is equal to the probability of just one overlap, which by Poisson statistics is then the expected number of overlaps. Accordingly, the total change in the number of sources at $z$ is then given by
\begin{align}
    \label{eq:so_dec_z}
    \Delta\, n^{\rm SO}(z, \boldsymbol{\theta}) &= - \pi\,\vartheta^2 \, n(z, \boldsymbol{\theta}) \int_0^{\infty} {\rm d}z' n(z', \boldsymbol{\theta}) \nonumber\\
    &=  -\pi\,\vartheta^2 \, n(z, \boldsymbol{\theta})\, [1+\delta^g(\boldsymbol{\theta})]\int_0^{\infty} {\rm d}z' n(z')\nonumber\\
    &=  -\pi\,\vartheta^2 \, n(z, \boldsymbol{\theta})\, [1+\delta^g(\boldsymbol{\theta})]\,n_{\rm tot}\, .
\end{align}

We neglect the second term on the right-hand side of Eq.~(\ref{eq:so_ndelt_full}) as it specifically accounts for the correlated overlap of galaxies at the same redshift within a fixed disk around the source, in addition to the random one already included. Given that the fractional change is calculated by integrating over redshift slices, that this term would only appear for the slice where $z'=z$, and that the source obscuration term itself is small (see Sect. \ref{sec:4}), we expect it to be safely negligible.

Adopting the tomographic redshift binning approach, for a given source can be obscured by another in the same bin, or by sources in lower redshift bins than the one the source belongs in. The fractional change in the number density of galaxies in redshift bin $i$ then becomes
\begin{align}
    \label{eq:so_ndelt}
    \frac{\Delta \,n^{\rm SO}_i(z, \boldsymbol{\theta})}{n_i(z, \boldsymbol{\theta})}&=-\pi\,\vartheta^2\sum_{q=1}^i [1+\delta^g_q(\boldsymbol{\theta})]\,n_{{\rm tot};\, q}\nonumber\\
    &= -\pi\,\vartheta^2\sum_{q=1}^i [1+b_q\delta^g(\boldsymbol{\theta})]\,n_{{\rm tot};\, q}\nonumber\\
    &= -\pi\,\vartheta^2\,n_{\rm cumul.; i} - \pi\,\vartheta^2\,\delta^g f_{{\rm SO; i}}\, ,
\end{align}
where the $n_{{\rm tot};\, q}$ is the total surface density of galaxies for redshift bin $q$ (that is the integral over $n_{q}(z)$ for bin $q$), $n_{\rm cumul.; i}$ is the cumulative total surface density of galaxies for all redshift bins up to and including bin $i$, and
\begin{align}
    \label{eq:fSOi}
    f_{{\rm SO; i}} = \sum_{\rho=1}^i b_\rho\, n_{{\rm tot};\, \rho}\, .
\end{align}

Including the effect of source obscuration in addition to source-lens clustering and magnification bias, the observed number density for a given tomographic redshift bin $i$, becomes

\begin{align}
    \label{eq:n_so_all}
    n_i^{\rm obs}(\boldsymbol{\theta}, \chi) &= n_i(\chi)\,\big[1 + (5s_i-2)\,\kappa_i(\boldsymbol{\theta}) + \delta^g_i(\boldsymbol{\theta})\,\big] \nonumber\\
    &\times \big[1-\pi\,\vartheta^2n_{\rm cumul.; i}-\pi\,\vartheta^2\delta^g(\boldsymbol{\theta})f_{{\rm SO}; i}\,\big]\, .
\end{align}
From here, only terms to first-order in the lensing potential are retained in order to suppress fourth-order or higher terms appearing in the two-point statistic. Then, it can be seen that source obscuration adds prefactors to the base angular shear power spectra, the source-lens clustering correction from Eq.~(\ref{eq:dCl_slc}), and the magnification bias correction of Eq.~(\ref{eq:dCl_MB}). Accordingly, source obscuration produces three new correction terms:
\begin{align}
    \label{eq:dCl_so}
    \delta C^{\rm SO}_{\ell;ij} &= \big(\pi^2\,\vartheta^4n_{\rm cumul.; i}n_{\rm cumul.; j} -\pi\,\vartheta^2n_{\rm cumul.; i}\nonumber\\
    &-\pi\,\vartheta^2n_{\rm cumul.; j}\big)\,C_{\ell;ij}^{\gamma\gamma}\, , \\
    \label{eq:dCl_soslc}
    \delta C^{\rm SO-SLC}_{\ell;ij} &= \int_0^\infty\frac{{\rm d}^2\boldsymbol{\ell}'}{(2\pi)^2}\cos(2\phi_{\ell'})\,\nonumber\\
    &\times \bigg[\big(\pi^2\,\vartheta^4n_{\rm cumul.; i}n_{\rm cumul.; j}-\pi\,\vartheta^2n_{\rm cumul.; i}b_i \nonumber\\
    &- \pi\,\vartheta^2n_{\rm cumul.; j}b_i -\pi\,\vartheta^2(1-n_{\rm cumul.; j})\nonumber\\
    &\times f_{{\rm SO}; i}\big)\,B_{iij}^{\kappa\delta^g \kappa}(\boldsymbol{\ell}_1, \boldsymbol{\ell}_2, \boldsymbol{\ell}_3)  + i \leftrightarrow j\bigg], \\
    \label{eq:dCl_somb}
    \delta C^{\rm SO-MB}_{\ell;ij} &= \big(\pi^2\,\vartheta^4n_{\rm cumul.; i}n_{\rm cumul.; j} -\pi\,\vartheta^2n_{\rm cumul.; i}\nonumber\\
    &-\pi\,\vartheta^2n_{\rm cumul.; j}\big)\,\delta C^{\rm MB}_{\ell;ij}\, ,
\end{align}
where $i \leftrightarrow j$ indicates a repetition of the preceding bispectrum term and its pre-factor, with all instances of the $i$ and $j$ bin indices exchanged.

\subsection{\label{subsec:localuniverse}Local Universe effects}

A further effect to consider is that the observed two-point statistic at our location may be biased due to local over or under-densities. Accordingly, the angular power spectra must be calculated conditioned on the local density \citep{LUHall}. The local density contrast, $\delta_{0}$, can be defined as the matter density contrast smoothed by a top-hat kernel of comoving radius $R$ according to
\begin{align}
    \label{eq:local_dens}
    \delta_{0}(R, \chi) \equiv \frac{3}{4 \pi R^{3}} \int \mathrm{d}^{3} \boldsymbol{r} \,\Theta(R-|\boldsymbol{r}|) \, \delta(\boldsymbol{r}, \chi)\, ,
\end{align}
where the matter density contrast is now expressed in terms of spatial distance, $\boldsymbol{r}$, rather than angle on the sky, and $\Theta$ is the Heaviside step-function.

Then, the conditional angular power spectra can be obtained using the Edgeworth expansion for conditional distributions. This calculation is mathematically intensive, and accordingly is not reproduced here. The full derivation can be found in \cite{LUHall}. Under the Limber approximation, and assuming $\ell \gg 1$ (and note that cosmic shear is only defined for $\ell\geq 2$), this expression consists of two terms, the standard power spectra of Eq.~(\ref{eq:Cl}) and a correction term, $\delta C^{\rm LU}_{\ell;ij}$ which is defined as
\begin{align}
    \label{eq:dCl_LU}
\delta C^{\rm LU}_{\ell;ij} &= 2 \frac{\delta_{0}(R, \chi)}{\sigma^{2}(R, \chi)} \int_0^{\chi_{\rm h}}{\rm d}\chi\,\frac{W_i(\chi)W_j(\chi)}{\chi^2}\nonumber\\
& \times\Bigg\{\left[\frac{34}{21} \xi_{R}(\chi)-\frac{4}{21} \psi_{R}(\chi)\right] P_{\delta\delta}(k, \chi)\nonumber\\
&+\left[\frac{\chi}{\ell} \xi_{R}^{\prime}(\chi)-\frac{\ell}{\chi} \Omega_{R}(\chi)\right] \frac{\partial P_{\delta\delta}(k, \chi)}{\partial k}\frac{1}{\chi}\nonumber\\
&-\frac{4}{7} \psi_{R}(\chi) \frac{\partial^2 P_{\delta\delta}(k, \chi)}{\partial k^2}\frac{1}{\chi^2}\Bigg\}\, ,
\end{align}
where $\sigma^2$ is the variance of the local density contrast, it is assumed the ratio of the local density contrast to its variance is constant with comoving distance, $\xi_{R}, \psi_{R}, \xi_{R}^{\prime}$, and $\Omega_{R}$ are correlation functions defined in \cite{LUHall}, and the expression assumes a flat-geometry, which is valid under current constraints on $\Omega_K$, as lenses are much less than curvature distance away. Additionally, we note that this expression is derived using only the tree-level Eulerian perturbation theory expression for the matter bispectrum.

\subsection{\label{subsec:flatgeo}The flat Universe assumption}

Typically when computing cosmic shear angular power spectra, spatially non-flat universes are accounted for through modifying comoving distances based on curvature, as described by Eq.~(\ref{eq:SK}). In practice, however, curvature also modifies the projection kernel \citep{Spatflatpap}.

Under the assumption of a spatially flat Universe, the Poisson equation gives the relationship between the comoving Newtonian gravitational potential, $\phi$, and the matter density contrast,
\begin{align}
    \label{eq:poisson}
    \nabla_{\chi}^{2}\, \phi(\boldsymbol{r}, \chi)=\frac{3 \Omega_{\rm m} H_{0}^{2}}{2\,c^2a(t)}\, \delta(\boldsymbol{r}, \chi)\, ,
\end{align}
where $\nabla_{\chi}^{2}$ is the Laplacian for a spatially flat Universe. This allows the shear angular power spectra to be expressed in terms of the matter power spectrum, as in Eq.~(\ref{eq:Cl}). However, the matter density contrast has rectilinear coordinates, whereas the lensing potential is defined in terms of angular coordinates $(r, \theta, \varphi)$, from the observer's frame of reference. Relating the two as above requires expressing the potential in spherical Bessel space as
\begin{align}
    \label{eq:sphpotential}
    \phi_{\ell m}(k)=\sqrt{\frac{2}{\pi}} \int \mathrm{d}^{3} r \, \phi(r) j_{\ell}(k r) Y_{\ell m}(\theta, \varphi)\, ,
\end{align}
where $j_{\ell}$ are spherical Bessel functions, and $Y_{\ell m}$ are spherical harmonics. Owing to the fact that these spherical Bessel functions and spherical harmonics are eigenfunctions of the Laplacian, the following relationship is obtained:
\begin{align}
    \label{eq:flat0eq}
    \left(\nabla_{r}^{2}+k^{2}\right) j_{\ell}(k r) Y_{\ell m}(\theta, \varphi)=0\, .
\end{align}
This allows the relation of the lensing potential to the matter density contrast in spherical harmonic space and, under the Limber approximation, the eventual calculation of Eq.~(\ref{eq:Cl}). See \cite{limitsofshear17} for a full derivation.

However, in the case of a spatially non-flat Universe, the Laplacian in Eqs. (\ref{eq:poisson} -- \ref{eq:flat0eq}) must be replaced by one corresponding to a curved geometry, $\nabla_{S_K}^{2}$. Accordingly, the projection kernel must also be modified; by replacing the spherical Bessel functions in Eq.~(\ref{eq:flat0eq}) with hyper-spherical Bessel functions, $\Phi_{\ell}^{\beta}$, so that
\begin{align}
    \label{eq:curv0eq}
    \left(\nabla_{S_K}^{2}+k^{2}\right) \, \Phi_{\ell}^{\beta}(r)\, Y_{\ell m}(\theta, \varphi)=0\, ,
\end{align}
where $\beta=\sqrt{(k^2 + K)\, /\, |K|}$ \citep{curvprefacpap}. Consequently, the shear angular power spectra for a spatially non-flat Universe, $C_{\ell;ij}^{\gamma\gamma; \rm {NF}}$, under the Limber approximation,under the Limber approximation, is given by modifying equation (\ref{eq:Cl}) to be 
\begin{eqnarray}
    C_{\ell;ij}^{\gamma\gamma} &=& \int_0^{\chi_{\rm h}}{\rm d}\chi W^{\rm NF}_{\ell}(\chi; K)\frac{W_i(\chi)W_j(\chi)}{S^{\,2}_K(\chi)}P_{\delta\delta}(k, \chi)\, ,
\end{eqnarray}
where 
\begin{eqnarray}
W^{\rm NF}_{\ell}(\chi; K)=\left[1 - {\rm sgn}(K)\frac{\ell^2}{(\ell+1/2)^2/S^2_K(\chi)+K}\right]^{-1/2}
\end{eqnarray}
where ${\rm sgn}(K)$ is the sign of the curvature $K$. Alternatively, for consistency with the previously discussed corrections, this can be expressed as a correction term, $\delta C_{\ell;ij}^{\rm NF}$, to the spatially flat angular power spectra such that
\begin{align}
    \label{eq:dCl_NF}
    \delta C_{\ell;ij}^{\rm NF} = C_{\ell;ij}^{\gamma\gamma; \rm {NF}} - C_{\ell;ij}^{\gamma\gamma}\,.
\end{align}

\subsection{\label{subsec:fisher}The Fisher matrix and bias formalism}
The constraining power of cosmological surveys, in terms of uncertainties on inferred cosmological parameters, is often predicted by using the Fisher matrix formalism. It also allows the quantification of biases in this inference resulting from neglecting systematic effects within the signal itself. Here, we use this technique to predict how biased cosmological parameters inferred from \emph{Euclid} would be when the previously discussed systematic effects are neglected.

Explicitly, the Fisher matrix is defined as the expected value of the Hessian of the log likelihood  \citep[defined for a Gaussian likelihood, and applied to CMB data in][]{Tegmark97}. For Stage IV weak lensing cosmology, it has been demonstrated that the likelihood can safely be assumed to be Gaussian \citep{LSSTnonGauss, TaylorNG, UphamGauss, HT22}. Accordingly, the Fisher matrix for cosmic shear is defined as
\begin{equation}
    \label{eq:fishGauss}
    F_{\mu\nu} = \sum_{\ell'=\ell_{\rm min}}^{\ell_{\rm max}}\sum_{\ell=\ell_{\rm min}}^{\ell_{\rm max}}\sum_{ij,mn}
             \frac{\partial C^{\epsilon\epsilon}_{\ell;ij}}{\partial \theta_{\mu}} {\rm Cov}^{-1}\left[C^{\
\epsilon\epsilon}_{\ell;ij},C^{\epsilon\epsilon}_{\ell';mn}\right]
\frac{\partial C^{\epsilon\epsilon}_{\ell';mn}}{\partial \theta_{\nu}}\, ,
\end{equation}
where the $\mu$ and $\nu$ indices denote element in the Fisher matrix associated with cosmological parameters $\theta_{\mu}$ and $\theta_{\nu}$ respectively, $\ell_{\rm min}$ is the minimum angular wavenumber of the survey, $\ell_{\rm max}$ is the maximum angular wavenumber used, the sums are over the $\ell$-blocks of power spectrum bands, and ${\rm Cov}^{-1}\left[C^{\
\epsilon\epsilon}_{\ell;ij},C^{\epsilon\epsilon}_{\ell';mn}\right]$ is the inverse of the covariance of the angular power spectra signal.

In practice, this covariance term is non-Gaussian \citep{WLSSCpap, HuSSC, UphamNonGauss}, with an additional contribution arising from the super-sample covariance \citep[SSC; ][]{SSCorig}. This SSC terms encapsulates the effects on the covariance of density fluctuations with wavelengths larger than the extent of the galaxy survey. Such fluctuations result in the background density of the survey ceasing to be representative of the underlying density of the Universe. The total covariance is then the sum of the Gaussian, ${\rm Cov_G}$ and SSC, ${\rm Cov_{SSC}}$, terms:
\begin{align}
    \label{eq:covsum}
    {\rm Cov}\left[C^{\epsilon\epsilon}_{\ell;ij},C^{\epsilon\epsilon}_{\ell^\prime;mn}\right] &= {\rm Cov}_{\rm G}\left[C^{\epsilon\epsilon}_{\ell;ij},C^{\epsilon\epsilon}_{\ell^\prime;mn}\right]\nonumber\\
    &+ {\rm Cov}_{\rm SSC}\left[C^{\epsilon\epsilon}_{\ell;ij},C^{\epsilon\epsilon}_{\ell^\prime;mn}\right]\, ,
\end{align}
where the Gaussian component is given by
\begin{align}
    \label{eq:gauscov}
    {\rm Cov_G}\left[C^{\epsilon\epsilon}_{\ell;ij},C^{\epsilon\epsilon}_{\ell^\prime;mn}\right] &=
\frac{C^{\epsilon\epsilon}_{\ell;i m}\,C^{\epsilon\epsilon}_{\ell^\prime;jn}+C^{\epsilon\epsilon}_{\ell;in}\,C^{\
\epsilon\epsilon}_{\ell^\prime;jm}}{(2 \ell + 1) f_{\rm sky} \Delta \ell}\,\delta^{\rm K}_{\ell\ell^\prime}\, ,
\end{align}
where $f_{\rm sky}$ is the fraction of the sky observed by the galaxy survey, $\Delta\ell$ is the bandwidth of the $\ell$-modes sampled, and $\delta^{\rm K}$ is the Kronecker delta. Other non-Gaussian terms in the covariance can be neglected \citep[see e.g.][]{2018JCAP...06..015B}. The SSC component is well-approximated by \citep{fastssc}
\begin{equation}
    \label{eq:covssc}
    {\rm Cov}_{\rm SSC}\left[C^{\epsilon\epsilon}_{\ell;ij},C^{\epsilon\epsilon}_{\ell^\prime;mn}\right] \approx R_\ell\, C^{\epsilon\epsilon}_{\ell;ij}\, R_{\ell'}\, C^{\epsilon\epsilon}_{\ell^\prime;mn}\,S_{ijmn}\, ,
\end{equation}
where $S_{ijmn}$ is the dimensionless volume-averaged covariance of the background matter density contrast, and $R_\ell$ is the effective relative response of the observed power spectrum. We assume that there is no interrelation between local Universe effects and the SSC, but this is a caveat that should be verified in future. 

The diagonal of the inverse of the Fisher matrix is used to predict the 1$\sigma$ uncertainties on each of the parameters. Explicitly, the uncertainty, $\sigma_\mu$, on parameter $\theta_\mu$ is given by 
\begin{align}
    \label{eq:sigfish}
    \sigma_\mu = \sqrt{{F_{\mu\mu}}^{-1}}\, .
\end{align}

By extending this formalism, the biases on inferred parameters resulting from neglecting systematic effects can also be predicted \citep{biaspap}. For a given systematic, $\delta C_{\ell; ij}$, the bias, $\mathfrak{b}_\mu$, on parameter $\theta_\mu$, is given by
\begin{align}
    \label{eq:biasfull}
    \mathfrak{b}_\mu &= \sum_\nu{(F^{-1})}_{\mu\nu}\: \mathcal{B}_\nu\, ,
\end{align}
where
\begin{align}
    \label{eq:biasextended}
    \mathcal{B}_{\nu} &=
\sum_{\ell'=\ell_{\rm min}}^{\ell_{\rm max}}\sum_{\ell=\ell_{\rm min}}^{\ell_{\rm max}}\sum_{ij,mn}\delta C_{\ell;ij}\,
{\rm Cov}^{-1}\left[C^{\
\epsilon\epsilon}_{\ell;ij},C^{\epsilon\epsilon}_{\ell';mn}\right]\;\frac{\partial C_{\ell';mn}}{\partial \theta_\nu}\, .
\end{align}

We note that we assume a Gaussian likelihood function but with a correlated  covariance matrix (that includes non-Gaussian contributions). The extent to which this assumption is robust to relaxing  the Gaussian likelihood assumption was explored in \cite{2021A&A...649A.100M} who found good agreement between Fisher matrix (Gaussian likelihood) predictions and full MCMC predictions, and in \cite{2019PhRvD.100b3519T} who found a similar result but allowing for the possibility of a fully non-Gaussian likelihood function.

\section{\label{sec:3}Methodology}

In this section, we review the computational and modelling specifics used within this investigation. We begin by describing the survey specifications adopted. Then, details are given about our choice of fiducial cosmology, modelling of background quantities, and Fisher matrices. Lastly, we describe modelling choices made in the computation of the magnification bias, source obscuration, and local Universe effects corrections.

\subsection{\label{subsec:surveyspec}Survey specifications}

\begin{table}[b]
    \centering
    \caption{Values of model parameters used in defining the uncertainty of photometric redshift estimates through Eq.~(\ref{eq:pphot}). Chosen according to EC20.}
    \begin{tabular}{c c}
    \hline\hline
    Parameter& Fiducial Value\\
    \hline
    $c_{\rm b}$ & 1.0\\
    $z_{\rm b}$ & 0.0\\
    $\sigma_{\rm b}$ & 0.05\\
    $c_{\rm o}$ & 1.0\\
    $z_{\rm o}$ & 0.1\\
    $\sigma_{\rm o}$ & 0.05\\
    $f_{\rm out}$ & 0.1\\
    \hline\hline
    \end{tabular}
    \label{tab:phphotparams}
\end{table}

For \emph{Euclid}, forecasting specifications are specified in EC20; we adopted these here, but we note that our IA model is more simple (see Section \ref{subsec:IAshotnoise}), and we vary $\Omega_K$ rather than dark energy critical density. Specifically, we considered the `optimistic' scenario described in that work, as this is the case where the cosmic shear probe is able to meet its precision goals by itself. Under this scenario, the survey is taken to extend up to $\ell$-modes of 5000.

Additionally, the intrinsic variance of observed ellipticities is taken to consist of two components; each with a magnitude of 0.21. Correspondingly, the RMS intrinsic ellipticity variance is $\sigma_\epsilon = \sqrt{2}\times0.21 \approx 0.3$ \footnote{We use the specification in EC20, but note that \cite{2019A&A...627A..59E} uses a value of $0.26$ per component. Since we are looking at biases caused by the differences in the signal the  shot noise component does not affect the reported biases, however the relative significance (bias divided by error) will be lower for a larger shot noise term.}. \emph{Euclid} is also expected to have a survey area such that $f_{\rm sky}=0.36$. The survey's galaxy surface density is anticipated to be $\bar{n}_{\rm g}=30$ arcmin$^{-2}$.

The cosmic shear probe of \emph{Euclid} is planned to observe sources between redshifts of 0 and 2.5, and utilise 10 equi-populated tomographic redshift bins with the following edges: \{0.001, 0.418, 0.560, 0.678, 0.789, 0.900, 1.019, 1.155, 1.324, 1.576, 2.50\}.

Given that \emph{Euclid} will use photometric redshifts, the model for the source distributions within these tomographic bins must account for photometric redshift uncertainties. Accordingly, for a particular bin, $i$, the galaxy redshift distribution, $n_i(z)$, was described by
\begin{equation}
    \label{eq:nphotoz}
    n_i(z) = \frac{\int_{z_i^-}^{z_i^+}{\rm d}z_{\rm p}\,\mathfrak{n}(z)p_{\rm ph}(z_{\rm p}|z)}{\int_{z_{\rm min}}^{z_{\rm max}}{\rm d}z\int_{z_i^-}^{z_i^+}{\rm d}z_{\rm p}\,\mathfrak{n}(z)p_{\rm ph}(z_{\rm p}|z)}\, ,
\end{equation}
where $z_{\rm p}$ is measured photometric redshift, $z_i^-$ and $z_i^+$ are the limits of the $i$-th redshift bin, and $z_{\rm min}$ and $z_{\rm max}$ are the redshift limits of the survey itself. Additionally, $\mathfrak{n}(z)$ is the underlying distribution of galaxies which here we modeled according to the formalism established in \cite{EuclidRB}:
\begin{equation}
    \label{eq:ntrue}
    \mathfrak{n}(z) \propto \bigg(\frac{z}{z_0}\bigg)^2\,{\rm exp}\bigg[-\bigg(\frac{z}{z_0}\bigg)^{3/2}\bigg]\, ,
\end{equation}
where $z_0=z_{\rm m}/\sqrt{2}$, and $z_{\rm m}=0.9$ is the median redshift of the survey. The remaining function in Eq.~(\ref{eq:nphotoz}), $p_{\rm ph}(z_{\rm p}|z)$, encapsulates the probability that a source measured to have a photometric redshift of $z_{\rm p}$ actually has a redshift of $z$. This distribution takes the form \citep{2008MNRAS.389..173K}
\begin{align}
\label{eq:pphot}
        p_{\rm ph}(z_{\rm p}|z) &= \frac{1-f_{\rm out}}{\sqrt{2\pi}\sigma_{\rm b}(1+z)}\,{\rm exp}\Bigg\{-\frac{1}{2}\bigg[\frac{z-c_{\rm b}z_{\rm p}-z_{\rm b}}{\sigma_{\rm b}(1+z)}\bigg]^2\Bigg\} \nonumber\\
        &+ \frac{f_{\rm out}}{\sqrt{2\pi}\sigma_{\rm o}(1+z)}\:{\rm exp}\Bigg\{-\frac{1}{2}\bigg[\frac{z-c_{\rm o}z_{\rm p}-z_{\rm o}}{\sigma_{\rm o}(1+z)}\bigg]^2\Bigg\}\, .
\end{align}
Here, the distribution is expressed as the sum of two terms -- the first is the uncertainty resulting from multiplicative and additive bias in redshift determination for the fraction of sources with a well measured redshift, whilst the second represents the same, but for a fraction of catastrophic outliers in the sample, $f_{\rm out}$. The values used for the individual parameters in this parameterisation match the selection of EC20, and are stated in Table \ref{tab:phphotparams}, which are fixed throughout our analysis.

\subsection{\label{subsec:Fishermethod}Cosmological modelling and Fisher matrices}

Throughout this investigation, we considered the $\Lambda$CDM cosmological model and its extension: the $w_0w_a$CDM model, which also allows for varying dark energy pressure and a separately parameterised dark energy equation of state at early times. The $\Lambda$CDM model uses 7 parameters, which are defined thusly: the present-day total matter density parameter $\Omega_{\rm m}$, the present-day baryonic matter density parameter $\Omega_{\rm b}$, the dimensionless curvature parameter $\Omega_K=-K(c/H_0)^2$, the Hubble parameter $h=H_0/100$\, km\:s$^{-1}$\,Mpc$^{-1}$, the spectral index $n_{\rm s}$, the RMS value of density fluctuations on 8 $h^{-1}\,$Mpc scales $\sigma_8$, and massive neutrinos with a sum of masses $\sum m_\nu\ne 0$. The $w_0w_a$CDM model additionally adds in the present-day value of the dark energy equation of state $w_0$, and the high-redshift value of the dark energy equation of state $w_a$. Typically the present-day densities $\Omega_{i}$, $i \in \{{\rm m, b, K}\}$, are denoted with an additional subscript 0; we omit this here for brevity. Primarily, we are interested the $w_0w_a$CDM case when discussing corrections in this investigation, as a key goal of Stage IV surveys is exploring models of dark energy. However, when examining the cosmological parameter biases, we also present the $\Lambda$CDM case, for completeness.

\begin{table}[t]
\centering
\caption{$\Lambda$CDM and $w_0w_a$CDM cosmological parameter fiducial values used in this investigation. These values correspond to EC20. It should be noted that $\sum m_\nu\ne 0$ is assumed to be fixed, and uncertainties and biases are not calculated for it. Additionally, two possible values are provided for $\Omega_K$, because the non-zero value must be used when evaluating the non-flat Universe correction. This value is selected using the upper-bound of the \cite{Planck18} 1$\sigma$ uncertainty.}
\label{tab:cosmology}
\begin{tabular}{c c}
\hline\hline
Cosmological Parameter & Fiducial Value\\
\hline
$\Omega_{\rm m}$ & 0.32 \\
$\Omega_{\rm b}$ & 0.05 \\
$h$ & 0.67 \\
$n_{\rm s}$ & 0.96 \\
$\sigma_8$ & 0.816 \\
$w_0$ & $-1$ \\
$w_a$ & 0  \\
$\sum m_\nu$ (eV) & 0.06 \\
$\Omega_K$ & \{0, 0.05\} \\
\hline\hline
\end{tabular}
\end{table}

The specific values used for each of these parameters are also chosen for consistency with EC20, and are given in Table \ref{tab:cosmology}. As in EC20, the value of $\sum m_\nu\ne 0$ was treated as fixed, and we did not calculate uncertainties or biases for it. When computing biases for all corrections except for the non-flat Universe term, we set $\Omega_K$ to 0. Only when testing the significance of the additional non-flat Universe correction term was it set to 0.05. In this case, the value is selected as being the upper-limit of the \cite{Planck18} 1$\sigma$ constraint on the parameter.

In the cases when the non-flat Universe correction was not being evaluated, our Fisher matrices matched those of the \emph{Euclid} forecasting specification (EC20), and contained: $\Omega_{\rm m}, \Omega_{\rm b}, h, n_{\rm s}, \sigma_8$ and $\mathcal{A}_{\rm IA}$ for the $\Lambda$CDM case, and additionally $w_0$ and $w_a,$ for the $w_0w_a$CDM case. When the correction for spatially curvature needed to be tested, the Fisher matrices also included $\Omega_K$. It was not necessary to include any further nuisance parameters within the matrix, as EC20 showed that the inclusion of various different nuisance parameters (such as those modelling the non-linear part of the matter power spectrum) typically altered the forecasted uncertainties on cosmological parameters by less than $10\%$. The $S_{ijmn}$ were calculated using the publicly available \texttt{PySSC}\footnote{\url{https://github.com/fabienlacasa/PySSC}} code \citep{fastssc}, with an $R_\ell$ of 3.

To calculate the cosmological background quantities required for the investigation, including the matter power spectrum and growth factor, we used the \texttt{CAMB}\footnote{\url{https://camb.info/}} software package \citep{cambpap}. Additionally, we utilised the \texttt{Halofit} \citep{Takahashi12} implementation of the non-linear part of the power spectrum, and included additional corrections identified by \cite{2012MNRAS.420.2551B}. Where necessary, we additionally employed  \texttt{Astropy}\footnote{\url{http://www.astropy.org}} \citep{astropy1, astropy2} to compute cosmological distances. The NLA model IA parameters were set to $\mathcal{A}_{\rm IA}=1.72$ and $\mathcal{C}_{\rm IA}=0.0134$, again in accordance with EC20. The required partial derivatives required were computed numerically, using the procedure described in EC20. Throughout this work, all quantities were evaluated for 200 $\ell$-bands. The limits for these were logarithmically spaced, with an $\ell_{\rm min}$ of 10, and an $\ell_{\rm max}$ of 5000.

\subsection{\label{subsec:corrspecs}Modelling higher-order corrections}

To model the matter bispectrum required by the reduced shear, magnification bias, source-lens clustering, and source obscuration corrections, we used the \texttt{BiHalofit} model and code\footnote{\url{http://cosmo.phys.hirosaki-u.ac.jp/takahasi/codes_e.htm}} \citep{bihalofit}. This represents the matter bispectrum using one-halo and three-halo terms, which themselves are determined through fitting to N-body simulations.

For the magnification bias correction, we used the slope of the luminosity function as calculated from the fitting formula given in Appendix C of \cite{EuclidMagGCWL}. This is determined from the \emph{Euclid} Flagship simulation \citep{Flagshippaper} and for the limiting magnitude 24.5 of the VIS instrument (AB in the \emph{Euclid} VIS band \citep{VISpap}). Therefore, it provides the most \emph{Euclid} specific estimate of this quantity to-date. However, once the \emph{Euclid} survey is in-progress, we note that this quantity should be calculated directly from the observed data. We used a single value for the slope for each tomographic redshift bin. This value was calculated at the central redshift of the bin. The slopes for all bins, together with their central redshifts, can be found in Table \ref{tab:slopesbybin}. We note that the magnification bias from \cite{EuclidMagGCWL} was obtained for the $n(z)$ from \cite{2021A&A...655A..44E} but we use the \cite{ISTforecast} $n(z)$, however the effect of the small changes in the assumed $n(z)$ should be small, which is consistent with the small differences in the results between this paper and \cite{EucRSMB}.
\begin{table}[t]
    \centering
    \caption{The values of the slope of the luminosity function used in computing the magnification bias correction. These are calculated at the central redshift of each tomographic bin. The limiting magnitude is taken to be 24.5, and the slopes are calculated with a fitting function \citep{EuclidMagGCWL} determined from the \emph{Euclid} Flagship simulation.} 
    \begin{tabular}{c c c c}
    \hline\hline
    Bin $i$ & Central Redshift & Slope $s_i$ \\
    \hline
    1 & 0.2095 & 0.108 \\
    2 & 0.489 & 0.180 \\
    3 & 0.619 & 0.229 \\
    4 & 0.7335 & 0.279 \\
    5 & 0.8445 & 0.335 \\
    6 & 0.9595 & 0.400 \\
    7 & 1.087 & 0.480 \\
    8 & 1.2395 & 0.586 \\
    9 & 1.45 & 0.753 \\
    10 & 2.038 & 1.335 \\
    \hline\hline
    \end{tabular}
    \label{tab:slopesbybin}
\end{table}

In order to evaluate the source obscuration terms, we set the total number of observed galaxies to $2\times 10^9$, so that the total number of galaxies per redshift bin was $2\times 10^8$. We also took the mean galaxy radius to be $\vartheta = 0.32''$ ($1.55\times 10^{-6}$ rad). This value is the mean half-light radius of galaxies from the \emph{Euclid} Flagship mock \citep{EuclidFlagshipGalsize}.

To evaluate the impact of the local Universe correction, we used a smoothing scale of 120$\,h^{-1}\,$Mpc. This is the primary value used in \cite{LUHall}, as it is just large enough for the local overdensity to be linear, while still possessing full-sky spherical coverage within the  2M++ galaxy redshift catalogue \citep{2M++} used to measure the local overdensity. Accordingly, $\delta_{0}\left(R=120 \,h^{-1}\, \mathrm{Mpc}\right) = 0.045$ and $\delta_{0}\left(R=120 \,h^{-1}\, \mathrm{Mpc}\right) / \sigma\left(R=120 \,h^{-1}\, \mathrm{Mpc}\right) = 0.85$. Different choices of smoothing scale result in different values of the local overdensity $\delta_0$, with some choices consistent with zero. As the LU effect scales linearly with the local overdensity, we do not rule out the LU bias as being exactly zero. Furthermore, it might be expected that the LU effect is mostly subsumed within the SSC uncertainty for any choice of smoothing scale, since it is mostly affected by modes that are outside of the survey. Detailed investigation of these points is beyond the scope of the paper, and our intention here is merely to assess how much bias would result from a nominal amplitude for the local overdensity combined with a fiducial implementation of the SSC covariance.

In this work, we compared the magnitude of the studied correction terms to the Gaussian sample variance, $\Delta C_\ell/C_\ell$. This was calculated according to \cite{Kaiser92}, and took the form
\begin{align}
    \label{eq:samplevar}
    \Delta C_\ell/C_\ell = \sqrt{2}\left[f_{\mathrm{sky}}(2 \ell+1)\right]^{-1 / 2}\, .
\end{align}

\section{\label{sec:4}Results and discussion}

\begin{figure*}[t]
    \centering
    \includegraphics[width=1.0\linewidth]{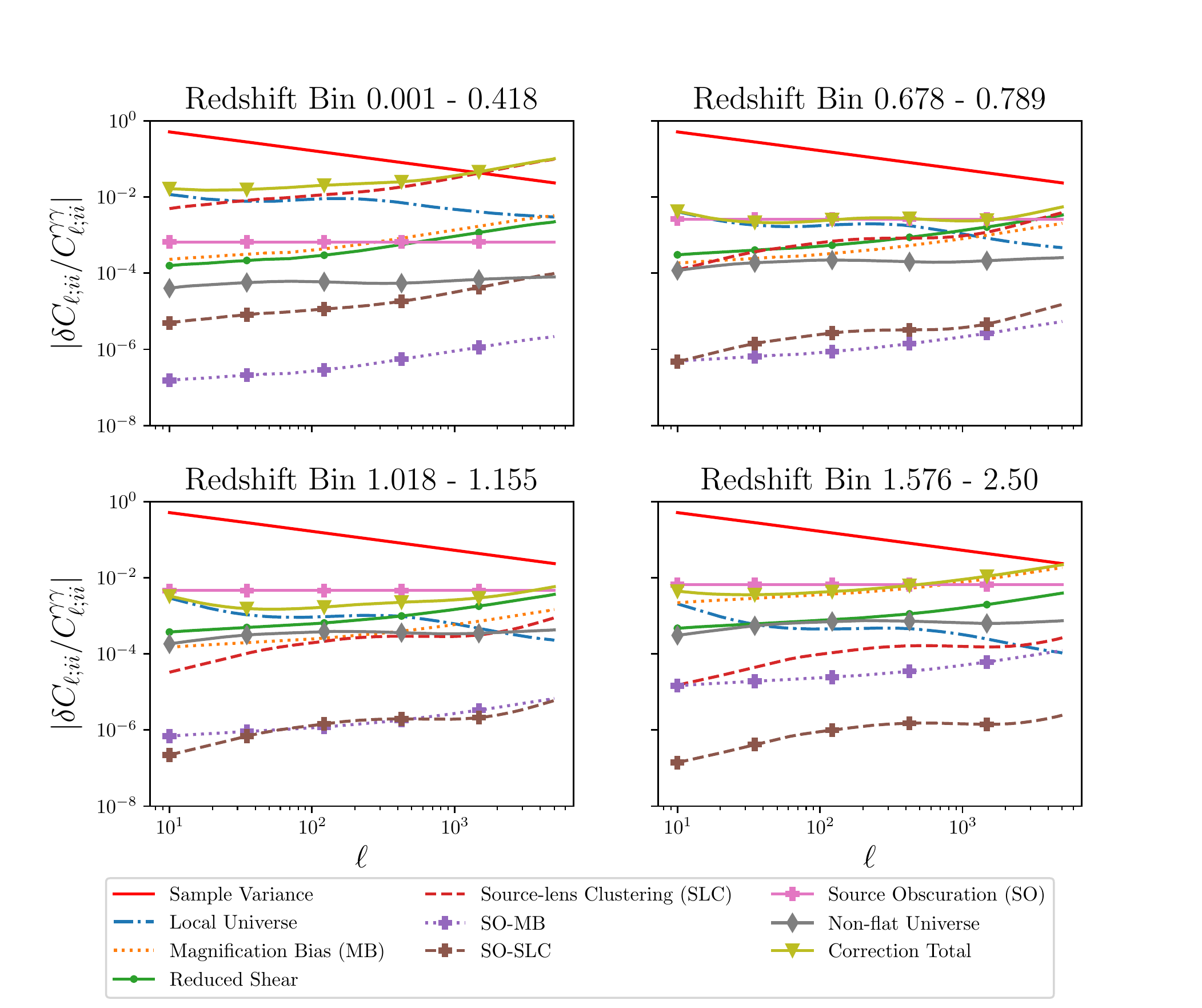}
    \caption{Absolute magnitudes of the reduced shear, source-lens clustering, magnification bias, local Universe, source obscuration, and non-flat Universe corrections to the shear angular power spectra, relative to those angular power spectra, for \emph{Euclid}. The corrections to the angular power spectra for four redshift bin auto-correlations are shown as representative examples, spanning across the redshift range of \emph{Euclid}. The remaining auto and cross-correlations exhibit the same patterns. The absolute value of the signed sum of the corrections is also shown. These are all compared to the sample variance, calculated according to Eq.~(\ref{eq:samplevar}). Notably, while the magnitudes of individual corrections are either higher at lower redshifts or vice-versa, the magnitude of the sum of the corrections is consistently high. Additionally, the cross-terms between source obscuration, and magnification bias and source-lens clustering are multiple orders of magnitude below other terms and sample variance, suggesting they are negligible. The remainder of the terms are typically of similar magnitudes across redshifts, suggesting they must all be accounted for. We note that these magnitudes are for both the $\Lambda$CDM and $w_0w_a$CDM cases, as the choice of fiducial values for the latter matches the former, and that the non-flat Universe correction here has been computed for a cosmology with $\Omega_{K}=0.05$, whilst other corrections are when $\Omega_{K}=0$. The markers for the SO-MB, SO-SLC, SO, Non-flat Universe, and total lines are only used to distinguish those from the other terms, and do not have any other significance. The symbols (points) are only included to allow a reader to distinguish the lines (in particular if printing in gray-scale) and do not indicate the $\ell$-modes where a computation was made; all quantities were evaluated for 200 $\ell$-bands, logarithmically spaced, with an $\ell_{\rm min}$ of 10, and an $\ell_{\rm max}$ of 5000.}
    \label{fig:dcls_all}
\end{figure*}

This section presents and discusses the computed values for the studied corrections. First, we show the magnitudes of the correction relative to the magnitude of the cosmic shear angular power spectra, and compare them to the sample variance.  

\begin{table*}[t]
\centering
\caption{The uncertainties on, and biases induced from neglecting the various corrections, in the $\Lambda$CDM parameters of Table \ref{tab:cosmology} for \emph{Euclid}. This table considers the case when $\Omega_{K}=0$, and accordingly lists the biases resulting from all corrections except the non-flat Universe correction. That can be found in Table \ref{tab:biasesnonflat}. All biases are given as a fraction of the $1\sigma$ uncertainty on each parameter. A bias is considered significant if it reaches or exceeds $0.25\sigma$, as at this point its uncorrected and corrected confidence contours overlap by less than $90\%$. `RS' denotes the reduced shear correction, `SLC' is the source-lens clustering term, `MB' is the magnification bias correction, `SO' is the two-point source obscuration correction, `SO-MB' and `SO-SLC' are the source obscuration-magnification bias and source-lens clustering cross terms respectively, and `LU' is the local Universe correction. The total biases from the sum of all corrections, as well as the total biases from only the individually significant corrections are also given.}
\label{tab:biasesLCDMtab}
\begin{tabular}{c c c c c c c c c c c }
\hline\hline
Cosmo. & Uncertainty & RS & SLC & MB & SO & SO-SLC & SO-MB & LU & Total & Total \\
Param. & $(1\sigma)$ & Bias/$\sigma$ & Bias/$\sigma$ & Bias/$\sigma$ & Bias/$\sigma$ & Bias/$\sigma$ & Bias/$\sigma$ & Bias/$\sigma$ & All & Sig. \\
\hline
$\Omega_{\rm m}$  & 0.0051 & $-$0.032  & 1.14    & $-$0.75 & 0.36                & 0.0032            & $0.0047$ & 0.26 & 0.76 & 0.73\\
$\Omega_{\rm b}$  & 0.021 & 0.0035 & 0.28    & 0.080 & $-0.044$ & $6.5\times 10^{-4}$ & $-6.6\times 10^{-4}$ & $-$0.016 & 0.26 & 0.28 \\
$h$               & 0.13  & 0.0084    & 0.15    & 0.19   & $-0.11$          & $2.8\times 10^{-4}$ & $-0.0013$ & $-$0.042 & 0.23 & 0.25\\
$n_{\rm s}$       & 0.029 & 0.017    & 0.0065 & $-$0.23 & 0.18            & $2.1\times 10^{-4}$  & $0.0016$ & 0.038 & $2.6\times 10^{-4}$ & $-0.07$\\
$\sigma_8$        & 0.0072 & 0.076     & $-$1.10 & 0.78    & $-$0.54           & $-0.0031$               & $-0.0050$ & $-$0.22 & $-$0.78 & $-0.79$\\
\hline\hline
\end{tabular}
\end{table*}

\begin{table*}[t]
\centering
\caption{The uncertainties on, and biases induced from neglecting the various corrections, in the $w_0w_a$CDM parameters of Table \ref{tab:cosmology} for \emph{Euclid}. This table considers the case when $\Omega_{K}=0$, and accordingly lists the biases resulting from all corrections except the non-flat Universe correction. That can be found in Table \ref{tab:biasesnonflat}. All biases are given as a fraction of the $1\sigma$ uncertainty on each parameter. A bias is considered significant if it reaches or exceeds $0.25\sigma$, as at this point its uncorrected and corrected confidence contours overlap by less than $90\%$. `RS' denotes the reduced shear correction, `SLC' is the source-lens clustering term, `MB' is the magnification bias correction, `SO' is the two-point source obscuration correction, `SO-MB' and `SO-SLC' are the source obscuration-magnification bias and source-lens clustering cross terms respectively, and `LU' is the local Universe correction. The total biases from the sum of all corrections, as well as the total biases from only the individually significant corrections are also given.}
\label{tab:biasesflattab}
\begin{tabular}{c c c c c c c c c c c }
\hline\hline
Cosmo. & Uncertainty & RS & SLC & MB & SO & SO-SLC & SO-MB & LU & Total & Total \\
Param. & $(1\sigma)$ & Bias/$\sigma$ & Bias/$\sigma$ & Bias/$\sigma$ & Bias/$\sigma$ & Bias/$\sigma$ & Bias/$\sigma$ & Bias/$\sigma$ & All & Sig. \\
\hline
$\Omega_{\rm m}$  & 0.010 & $-$0.079  & 1.25    & 0.12 & 0.12                & 0.0025            & $-0.0022$ & 0.14 & 1.55 & 1.49\\
$\Omega_{\rm b}$  & 0.021 & $-$0.0013 & 0.30    & 0.069 & $-0.021$ & $7.2\times 10^{-4}$ & $-5.8\times 10^{-4}$ & $-$0.0032 & 0.35 & 0.35 \\
$h$               & 0.13  & 0.0057    & 0.11    & 0.11   & $-0.053$            & $3.2\times 10^{-4}$ & $-6.3\times10^{-4}$ & $-$0.015 & 0.16 & 0.17\\
$n_{\rm s}$       & 0.031 & 0.051    & $-$0.19 & $-$0.19 & $0.026$            & $-3.0\times 10^{-4}$  & $0.0015$ & $-$0.045 & $-$0.35 & $-$0.35\\
$\sigma_8$        & 0.012 & 0.081     & $-$1.19 & 0.017    & $-0.19$           & $-0.0024$               & $0.0011$ & $-$0.093 & $-$1.38 & $-$1.36\\
$w_0$             & 0.13  & $-$0.076  & 0.81    & 0.55 & $-0.039$                & $0.0011$ & $-0.0051$ & 0.032 & 1.28 & 1.31\\
$w_a$             & 0.36  & 0.022     & $-$0.51 & $-$0.60    & 0.27            & $-3.3\times 10^{-4}$  & $0.0053$ & 0.097 & $-$0.72 & $-$0.84\\
\hline\hline
\end{tabular}
\end{table*}

\begin{table*}[t]
\centering
\caption{The uncertainties for the fiducial $\Lambda$CDM and $w_0w_a$CDM cosmologies of Table \ref{tab:cosmology} when $\Omega_{K}=0.05$, and biases induced from neglecting the non-flat Universe correction for the \emph{Euclid} cosmic shear probe. The biases are given as a fraction of the $1\sigma$ uncertainty on each parameter. A bias is considered significant if it reaches or exceeds $0.25\sigma$, as at this point its uncorrected and corrected confidence contours overlap by less than $90\%$. The predicted biases from this correction are well below significance for all parameters.}
\label{tab:biasesnonflat}
\begin{tabular}{c c c c c}
\hline\hline
Cosmo. & $\Lambda$CDM Uncertainty & $\Lambda$CDM Non-flat Universe & $w_0w_a$CDM Uncertainty & $w_0w_a$CDM Non-flat Universe \\
Param. & $(1\sigma)$ & Bias/$\sigma$ & $(1\sigma)$ & Bias/$\sigma$\\
\hline
$\Omega_{\rm m}$ &0.0050& 0.043 & 0.014 & $-$0.062\\
$\Omega_{\rm b}$ &0.021& $-$0.0070 & 0.021 & $8.1\times 10^{-4}$\\
$\Omega_{K}$     &0.034& 0.043 & 0.062 & $-$0.054\\
$h$              &0.12& $-$0.012 & 0.12  & 0.012\\
$n_{\rm s}$      &0.029& $-$0.010 & 0.030 & 0.012\\
$\sigma_8$       &0.028& 0.026 & 0.042 & $-$0.044\\
$w_0$            &N/A& N/A & 0.14  & $-$0.090\\
$w_a$            &N/A& N/A & 0.56  & 0.10\\
\hline\hline
\end{tabular}
\end{table*}

\begin{figure*}[t]
    \centering
    \includegraphics[width=1.0\linewidth]{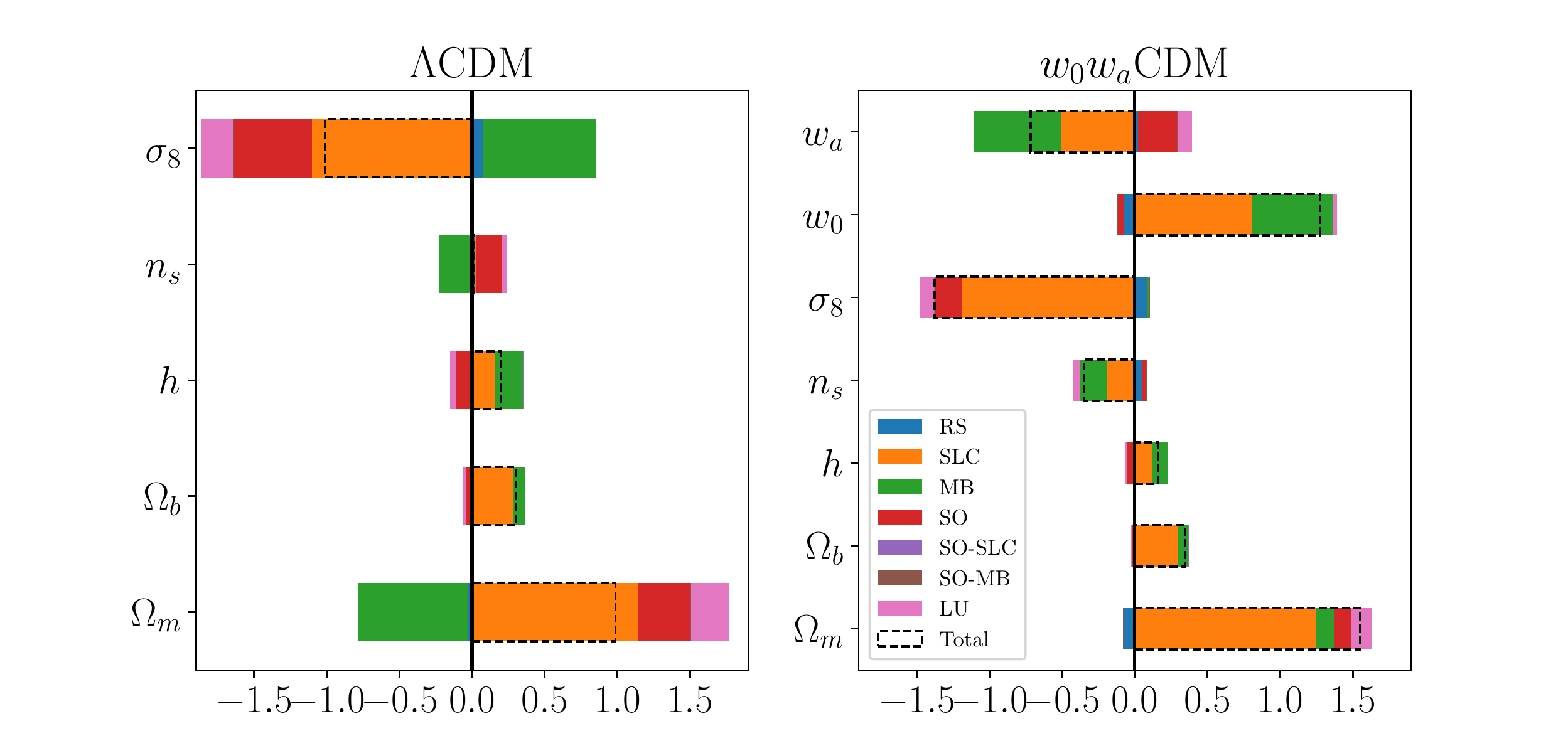}
    \caption{Stacked bar chart of cosmological parameter biases resulting from the studied higher-order effects, for the flat $\Lambda$CDM case (left) of Table \ref{tab:biasesLCDMtab}, and the flat $w_0w_a$CDM case (right) of Table \ref{tab:biasesflattab}. The non-flat Universe term is not shown here, due to the different cosmology. Biases are presented here as a fraction of the $1\sigma$ parameter uncertainty. A bias is non-negligble if its absolute value reaches or exceeds $0.25\sigma$. `RS' denotes the reduced shear correction, `SLC' is the source-lens clustering term, `MB' is the magnification bias correction, `SO' is the two-point source obscuration correction, `SO-MB' and `SO-SLC' are the source obscuration-magnification bias and source-lens clustering cross terms respectively, and `LU' is the local Universe correction The segments with the dashed outlines show the total parameter biases from these corrections for each parameter.}
    \label{fig:stacked_bars}
\end{figure*}

In Fig.~\ref{fig:dcls_all} the magnitudes of the reduced shear, source-lens clustering, magnification bias, source obscuration, local Universe effect, and non-flat Universe corrections, relative to the angular power spectra are shown. The combined correction is also shown, and we note that this is the sum of the signed values of the corrections, rather than the absolute values which are shown here for comparison. These terms are displayed for the auto-correlations of four redshift bins across the survey's range; specifically, bins 1, 4, 7, and 10. These particular bins are presented for illustrative purposes, and the remaining bins and cross-correlations display consistent trends. It can be seen that all corrections are typically below sample variance both individually and when combined, with the exception of at small physical scales at the lowest and highest redshifts. A noteworthy detail from this figure is that, while individual corrections are either consistently well below sample variance, or are higher at low redshifts and reduce significantly at high redshifts or vice-versa, the total magnitude of the corrections is consistently high. \begin{figure*}[t]
    \centering
    \includegraphics[width=1.0\linewidth]{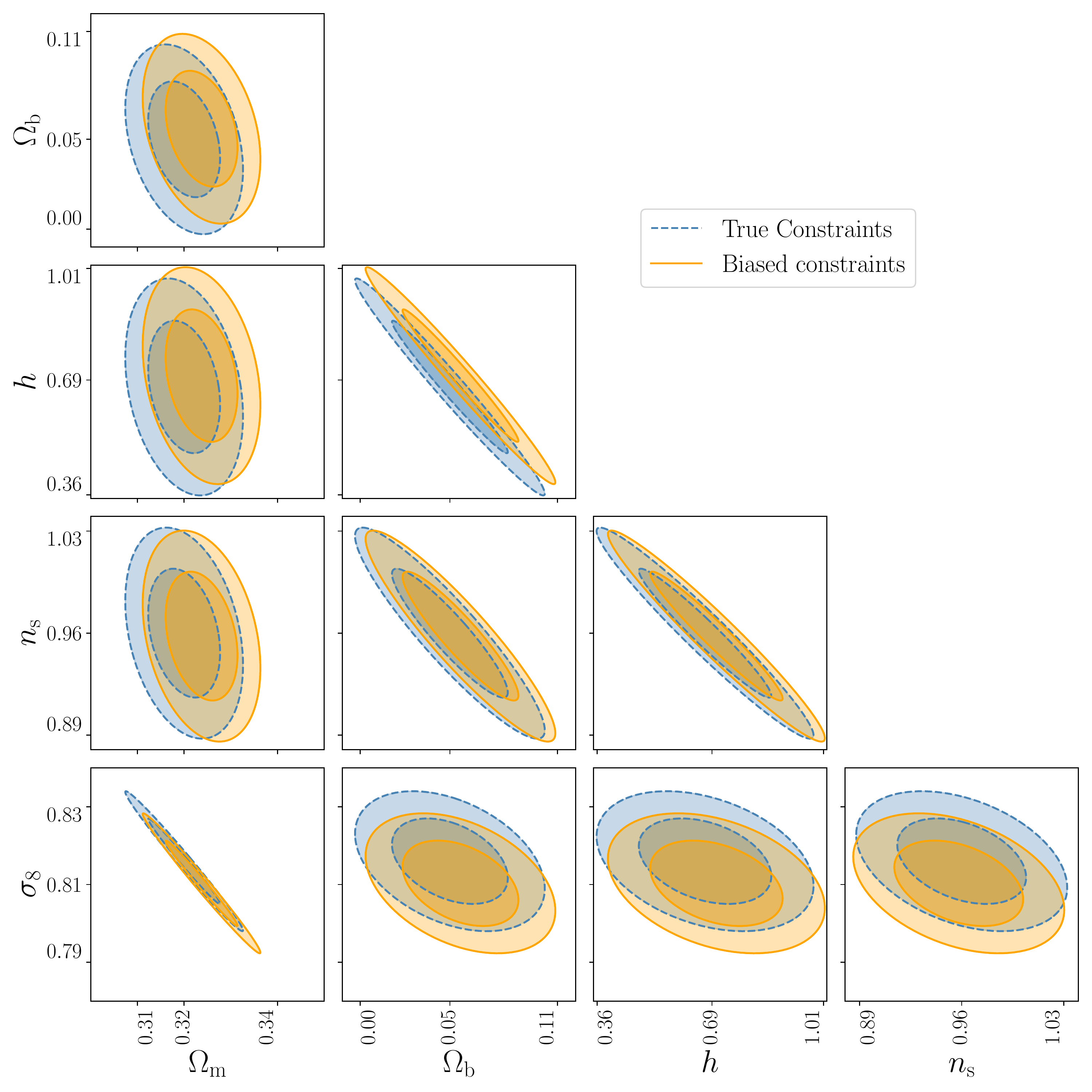}
    \caption{Projected 1$\sigma$ and 2$\sigma$ 2-parameter uncertainty contours for \emph{Euclid} under a $\Lambda$CDM cosmology, with and without correcting for the source-lens clustering, magnification bias, source obscuration, and local Universe terms. These are predicted using the Fisher matrix formalism, using the cosmology specified in Table \ref{tab:cosmology}, in the case when $\Omega_{K}=0$ and is kept fixed. The true location of the constraints is denoted by the blue, dashed contours, while the biased locations if the corrections are not made are given by the solid, gold contours. Significant biases are predicted for $\Omega_{\rm m}$, $\Omega_{\rm b}$, $h$, and $\sigma_{8}$, and their values can be found in Table \ref{tab:biasesLCDMtab}}
    \label{fig:bias_corner_LCDM}
\end{figure*}

Furthermore, another immediately noticeable feature is that the source obscuration cross terms with magnification bias and source-lens clustering are always multiple orders of magnitude smaller than sample variance, and typically the other terms as well; suggesting that these cross-terms are negligible.

Despite the fact that these terms are generally below sample variance, because they make contributions consistently across $\ell$-modes, they can still cause significant biases in inferred cosmological parameters. The bias in an estimated parameter resulting from neglecting a systematic effect is typically considered significant if it exceeds 25$\%$ of the $1\sigma$ uncertainty on that parameter \citep{biaspap}. This is because, at that point, the biased and unbiased 1$\sigma$ confidence contours overlap by less than 90$\%$.

\begin{figure*}[t]
    \centering
    \includegraphics[width=1.0\linewidth]{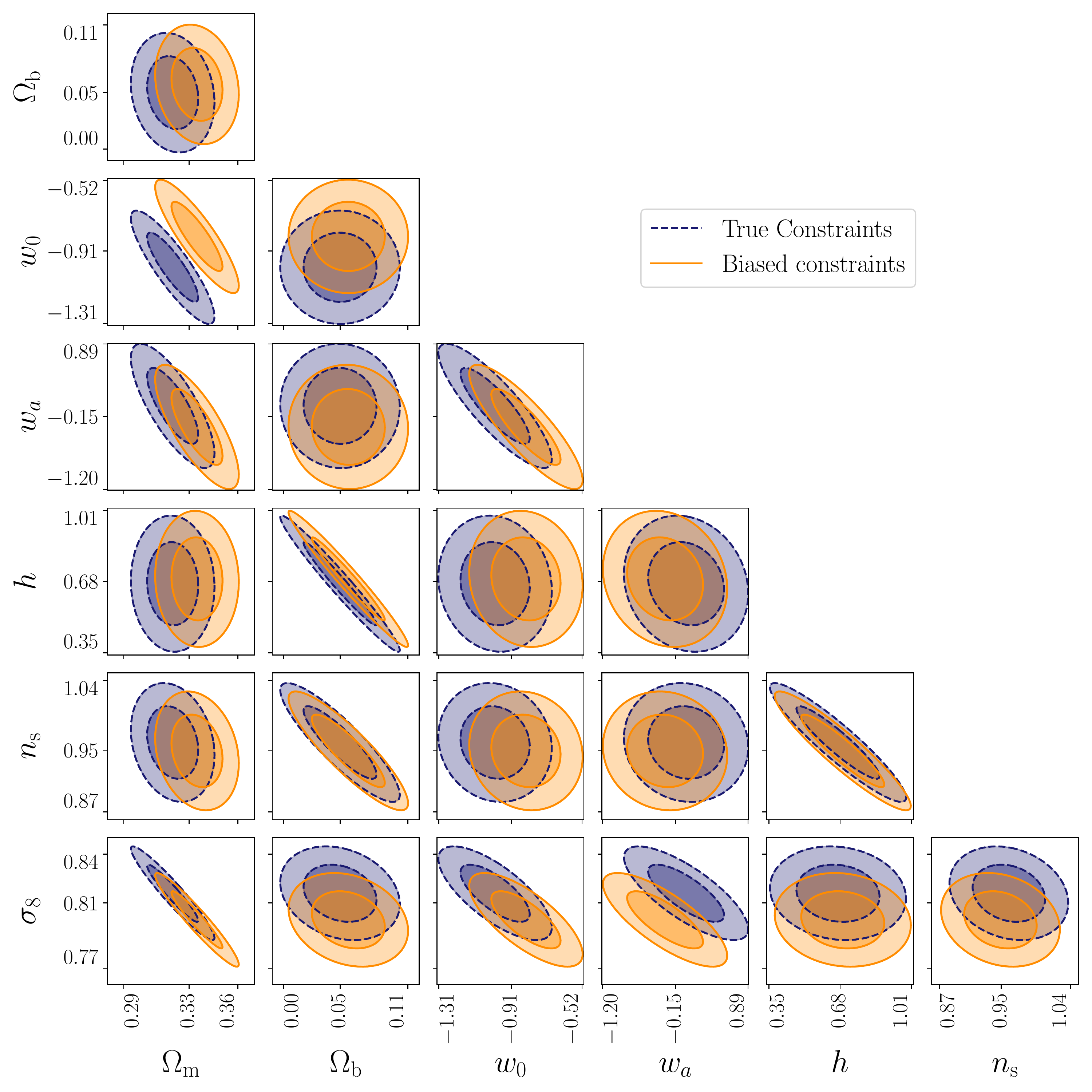}
    \caption{Projected 1$\sigma$ and 2$\sigma$ 2-parameter uncertainty contours for \emph{Euclid}, with and without correcting for source-lens clustering, magnification bias, and source obscuration. These are predicted using the Fisher matrix formalism, for the $w_0w_a$CDM cosmology specified in Table \ref{tab:cosmology}, in the case when $\Omega_{K}=0$ and is kept fixed. The true location of the constraints is denoted by the blue, dashed contours, while the biased locations if the corrections are not made are given by the solid, orange contours. Significant biases are predicted for $\Omega_{\rm m}$, $\Omega_{\rm b}$, $n_{\rm s}$, $\sigma_{8}$, $w_0$, and $w_a$, and their values can be found in Table \ref{tab:biasesflattab}.}
    \label{fig:bias_corner}
\end{figure*}

The predicted cosmological parameter biases resulting from neglecting all corrections except for the non-flat Universe term are stated in Table \ref{tab:biasesLCDMtab} and Table \ref{tab:biasesflattab}, for the $\Lambda$CDM and $w_0w_a$CDM cases, respectively. These are also represented visually in Fig.~\ref{fig:stacked_bars}. The biases from the non-flat Universe correction (which requires a fiducial cosmology with non-zero curvature) are stated in Table \ref{tab:biasesnonflat}, for both choices of cosmology. From these tables we see that the  source-lens clustering, magnification bias, source obscuration, and local Universe terms are individually significant in the $\Lambda$CDM case, while instead only the source-lens clustering, magnification bias, and source obscuration corrections are individually of concern in the $w_0w_a$CDM case. This difference is likely due to the presence of the variable dark energy parameters in the $w_ow_a$CDM scenario reducing sensitivity to scales where the local Universe term is important. Of these, the source-lens clustering term is particularly concerning, as for this term all but two of the parameters have significant biases.

Also shown in these tables are the combined biases when all of the individually significant corrections are taken into account, and when all of these corrections are taken into account. The full total is also shown in Fig.~\ref{fig:stacked_bars}. Owing to the fact that some biases are additive while others are subtractive, the total biases in the $\Lambda$CDM scenario are, in fact, less severe than some of the individual ones; in particular, source-lens clustering. However, the totals are still significant, and because they do not strongly resemble any one of the biases uniquely, multiple terms must still be computed. In the $w_0w_a$CDM case, the magnification bias no longer suppresses the source-lens clustering term, instead adding to it and meaning that the total biases in this case are more severe than the individual ones. This change likely occurs due to the dark energy terms increasing sensitivity to scales where the opposite component of the magnification bias (e.g. decrease in galaxy number density due to dilution rather than increase due to increased flux) is dominant.

At inference time, the computational load can be reduced by noting that only the individually significant terms (source-lens clustering, magnification bias, source obscuration, and local Universe) need to be computed in both cases, because this total does not significantly differ from the full total. Although, given that the reduced shear correction is also obtained at no additional cost when computing the magnification bias correction, we recommend including this too. The bias in the two-parameter confidence contours resulting from neglecting the combined effect of the significant biases is shown in Fig.~\ref{fig:bias_corner_LCDM} and Fig.~\ref{fig:bias_corner} for the $\Lambda$CDM and $w_0w_a$CDM scenarios, respectively. These figures display the contours in the case where no correction has been made, and when the corrections are made. As with Table \ref{tab:biasesLCDMtab} and Table \ref{tab:biasesflattab}, we see that the cumulative corrections must be accounted for.

We note that many of the modelling specifics used here, for example the slope of the luminosity function or the smoothing scale of local overdensity, will need to be determined directly from the \emph{Euclid} survey itself for self-consistency when these corrections are computed at inference time. Accordingly, it would not be meaningful to place constraints on them here, due to the variable survey specifics. 

Additionally, we note that the true value of the local density contrast, and accordingly the radius of the smoothing kernel used to calculate it, remain open questions. Accordingly, there is a large uncertainty in the local Universe correction which cannot be meaningfully constrained and it is possible that it may even be zero. Accurate measurements of the local density contrast are required for this.

Similarly, the source obscuration correction as computed in this work represents a worst-case scenario where \emph{every} galaxy in the foreground of a given redshift slice has an overlap. This represents an upper limit on the bias from this effect, but, in practice, the number of sources with overlap will be a smaller fraction. Accordingly, it is likely that source obscuration will not result in significant biases for true \emph{Euclid} observations, particularly if a robust mitigation strategy is employed.

However, the dominant source of quantifiable modelling uncertainty comes from the modelling of the matter bispectrum and, given that the bispectrum is not currently well constrained by observations, this model is likely to continue to evolve. Accordingly, it is important to constrain the impact of a change in bispectrum model on these terms. To-date, three widely-used matter bispectrum models have been produced \citep{Scoccimarro01, Gilmarinbispec, bihalofit}. As each subsequent model has become more complex and improved upon the accuracy of its predecessor, comparing correction magnitudes using each of these models would not realistically constrain the uncertainty from the bispectrum model.

Instead, it is useful to set a threshold around the latest of these models \citep{bihalofit}, within which any change in the model must be contained, in order to not produce a significant change in the correction terms. We do this by determining the minimum fractional increase or decrease in the matter bispectrum required across all triangle configurations, for each correction individually, to cause a significant change in any of the cosmological parameter biases, in the $w_0w_a$CDM case. This corresponds to a change of $\pm 0.25\sigma$ in any one of the biases. Given that we are placing multiplicative limits on the change in the bispectrum, for each correction the smallest limits are found when considering the cosmological parameter that already has the largest bias. The resulting limits are stated in Table \ref{tab:bispecfracchanges}, alongside the parameter which would see the corresponding significant change in its bias. From this, we see that the corrections most susceptible to a change in the bispectrum model are the source-lens clustering and magnification bias terms, as neglecting these already creates the most significant biases individually. Additionally, the two source obscuration cross terms are the least sensitive, requiring a change of an order-of-magnitude. This further reinforces the fact that these terms are safely negligible. We stress that these multiplicative limits are not exhaustive cut-offs on when a bispectrum model would cause a significant change, because a model with a sufficiently large change for only a select sub-range of scales or configurations could still cause a significant difference in the parameter biases. We recommend explicit revaluation with any future updated bispectrum models, should they non-trivially exceed these thresholds frequently.

Another consideration is how the inclusion of the studied effects would affect the size of the cosmological parameter uncertainty constraints themselves. In this work, we do not explicitly calculate this resulting change. However, it has previously been shown that for the bispectrum-dependent terms, even corrections that cause biases of greater than 1$\sigma$ result in negligible changes to the uncertainty constraints \citep{Shapiro09, EucRSMB}.

\section{\label{sec:5}Conclusions}

In this investigation, we have examined the higher-order corrections to the cosmic shear angular power spectra that must be modelled when performing inference with \emph{Euclid}. By first reviewing the literature, we identified 24 correction terms, and gathered representative values to facilitate comparison. From these, we identified six corrections which were potentially important for \emph{Euclid}, and evaluated them explicitly. These were: the relaxation of the reduced shear approximation, the source-lens clustering correction, the magnification bias correction, the source obscuration correction, the local Universe correction, and the non-flat Universe correction.

\begin{table}[t]
\centering
\caption{The increasing and decreasing multiplicative changes required in the matter bispectrum, across all configurations and all $\ell$-modes, to significantly affect the biases from each relevant correction term. A significant change is when any one of the biases on the cosmological parameters changes by $\pm 0.25\sigma$. These values encapsulate the fractional change for a fixed cosmology, in this case the $w_0w_a$CDM cosmology of Table \ref{tab:cosmology}, with $\Omega_K = 0$. It should be noted that these values are not exhaustive thresholds, as sufficiently significant changes to the model at a particular sub-range of scales may also be problematic.}
\label{tab:bispecfracchanges}
\begin{tabular}{c c c c}
\hline\hline
Correction & Min. Change & Bispec. Mult. & Bispec. Mult. \\
 & Param & Increase & Decrease \\
\hline
RS  & $\Omega_{\rm m}$ & $-$2.16 & $-$4.16\\
SLC & $\Omega_{\rm m}$ & 1.20 & 0.80\\
MB     & $w_a$ & 0.58 & 1.41\\
SO-SLC     & $\Omega_{\rm m}$ & 100.63 & $-99.67$\\
SO-MB      & $w_a$ & 47.00 & $-$47.00\\
\hline\hline
\end{tabular}
\end{table}

After calculating these corrections, we used the Fisher matrix formalism to predict the biases in cosmological parameter biases if each of these terms were to be neglected, in order to identify which ones are necessary to be modelled for \emph{Euclid}. This was done for two scenarios: a $\Lambda$CDM cosmology, and a $w_0w_a$CDM cosmology. For the first of these scenarios, we found that the source-lens clustering, magnification bias, source obscuration, and local Universe terms were significant, while for the second case we found that the source-lens clustering, magnification bias, and source obscuration corrections each produced significant biases in multiple parameters individually. The source-lens clustering term was noted as being of particular concern as multiple biases approached or exceeded $1\sigma$. However, in the $\Lambda$CDM case we found that when the biases are combined, they frequently suppressed each other, leading the total bias to be lower than many of the individual biases. Despite this, the total biases were still significant, and did not strongly represent the exact biases from any correction individually. In the $w_0w_a$CDM case, the total of the three biases was higher than the individual terms. Accordingly, we recommend that the source-lens clustering, magnification bias, source obscuration, and local Universe corrections are all taken into account when modelling the shear angular power spectra for \emph{Euclid}. Additionally, given that the reduced shear correction is obtained at no additional cost when computing the magnification bias correction, we recommend that this too be included. 

To provide some constraints on the predictive ability of this work, we quantified how much the matter bispectrum would have to change by in order to illicit a significant change in the biases predicted in this work. We identified that a $\sim20-40\%$ increase or decrease in the amplitude of the matter bispectrum at all scales, and for all triangle configurations would be required for this.

We note that this work did not investigate the impact of higher-order corrections on the IA spectra. Typically, even for effects which, when neglected, produce high biases, i.e. $\sim$O($1\sigma$), the corresponding corrections to the IA spectra cause negligible biases \citep{EucRSMB}. Accordingly, an explicit evaluation should not be necessary. Furthermore, we did not consider the impact of baryonic feedback on the bispectrum, and therefore on its dependant corrections. The impact of this remains poorly understood, with inconsistent findings from different simulations \citep{Owlsbisup, Illustrisbisup}. Accordingly this is out of the scope of this investigation. However, we note that given the magnitude of change required to the matter bispectrum in order to cause a significant change in the cosmological parameter biases, it is unlikely baryonic feedback would significantly alter the predictions of this investigation.

Given that we find it is necessary to include these higher-order terms in the modelling of the shear power spectra, an open question is the optimal strategy to do so. It has repeatedly been shown that computing these corrections for just one cosmology is relatively time consuming \citep{EucRSMB, MBSims}, rendering computation at inference time a serious challenge. While it may be possible to sufficiently optimise the required evaluation code, alternate strategies may also prove useful. Scale cutting techniques such as $k$-cut cosmic shear \citep{kcutpap} have been shown to mitigate the need to make such corrections without significantly compromising the constraining power of Stage IV surveys \citep{kcutrs}. Alternatively, emulation has recently become a popular tool in cosmology for reducing computation time at inference by replacing analytical models with emulators \citep[see e.g. recent work emulating the matter power spectrum within][]{cosmopower}. Emulators could also be developed directly for these correction terms, or intermediate quantities such as the matter or convergence bispectra.

Furthermore, while our analysis here is limited to the angular power spectra, significant corrections for this statistic are also likely to be significant for the two-point correlation function. In fact, due to the mode-mixing that occurs when transforming the power spectra to correlation functions, the effect of the discussed approximations is likely to be more severe. This, combined with the sensitivity of the correlation function to higher $\ell$-modes and the additional approximations required \citep[e.g. the flat Hankel transform; ][]{limitsofshear17}, means that if correlation functions were to be used a similar but separate study would be required to demonstrate modelling of the correlation function to higher-order corrections. 

\begin{acknowledgements}
ACD and AH are supported by the Royal Society. TDK acknowledges funding from the EU’s Horizon 2020 programme, grant agreement No 776247. The Euclid Consortium acknowledges the European Space Agency and a number of agencies and institutes that have supported the development of \emph{Euclid}, in
  particular
the Academy of Finland, the Agenzia Spaziale Italiana,
the Belgian Science Policy, the Canadian Euclid Consortium, the French Centre
National d'Etudes Spatiales, the Deutsches Zentrum f\"ur Luft- und
Raumfahrt, the Danish Space Research Institute, the Funda\c{c}\~{a}o
para a Ci\^{e}ncia e a Tecnologia, the Ministerio
de Ciencia e Innovaci\'on, the National Aeronautics and Space Administration, the
National Astronomical Observatory of Japan, the
Netherlandse Onderzoekschool Voor Astronomie, the Norwegian Space Agency, the Romanian Space Agency, the State Secretariat for Education, Research and Innovation (SERI) at the Swiss Space Office (SSO), and the United Kingdom Space Agency. CJM acknowledges FCT and POCH/FSE (EC) support through Investigador FCT 
Contract 2021.01214.CEECIND/CP1658/CT0001. A complete and detailed list is available on the \emph{Euclid}\ web site 
(\url{http://www.euclid-ec.org}). For the purpose of open access, the author has applied a Creative Commons Attribution (CC BY) licence to any Author Accepted Manuscript version arising from this submission.\xspace
\end{acknowledgements}


\bibliographystyle{aa}
\bibliography{aanda}

\providecommand{\noopsort}[1]{}\providecommand{\singleletter}[1]{#1}%
\begin{thebibliography}{96}
\expandafter\ifx\csname natexlab\endcsname\relax\def\natexlab#1{#1}\fi

\bibitem[{{Abbott} {et~al.}(2022){Abbott}, {Aguena}, {Alarcon}, {Allam},
  {Alves}, {Amon}, {Andrade-Oliveira}, {Annis}, {Avila}, {Bacon}, {Baxter},
  {Bechtol}, {Becker}, {Bernstein}, {Bhargava}, {Birrer}, {Blazek},
  {Brandao-Souza}, {Bridle}, {Brooks}, {Buckley-Geer}, {Burke}, {Camacho},
  {Campos}, {Carnero Rosell}, {Carrasco Kind}, {Carretero}, {Castander},
  {Cawthon}, {Chang}, {Chen}, {Chen}, {Choi}, {Conselice}, {Cordero},
  {Costanzi}, {Crocce}, {da Costa}, {da Silva Pereira}, {Davis}, {Davis}, {De
  Vicente}, {DeRose}, {Desai}, {Di Valentino}, {Diehl}, {Dietrich}, {Dodelson},
  {Doel}, {Doux}, {Drlica-Wagner}, {Eckert}, {Eifler}, {Elsner}, {Elvin-Poole},
  {Everett}, {Evrard}, {Fang}, {Farahi}, {Fernandez}, {Ferrero}, {Fert{\'e}},
  {Fosalba}, {Friedrich}, {Frieman}, {Garc{\'\i}a-Bellido}, {Gatti},
  {Gaztanaga}, {Gerdes}, {Giannantonio}, {Giannini}, {Gruen}, {Gruendl},
  {Gschwend}, {Gutierrez}, {Harrison}, {Hartley}, {Herner}, {Hinton},
  {Hollowood}, {Honscheid}, {Hoyle}, {Huff}, {Huterer}, {Jain}, {James},
  {Jarvis}, {Jeffrey}, {Jeltema}, {Kovacs}, {Krause}, {Kron}, {Kuehn},
  {Kuropatkin}, {Lahav}, {Leget}, {Lemos}, {Liddle}, {Lidman}, {Lima}, {Lin},
  {MacCrann}, {Maia}, {Marshall}, {Martini}, {McCullough}, {Melchior},
  {Mena-Fern{\'a}ndez}, {Menanteau}, {Miquel}, {Mohr}, {Morgan}, {Muir},
  {Myles}, {Nadathur}, {Navarro-Alsina}, {Nichol}, {Ogando}, {Omori},
  {Palmese}, {Pandey}, {Park}, {Paz-Chinch{\'o}n}, {Petravick}, {Pieres},
  {Plazas Malag{\'o}n}, {Porredon}, {Prat}, {Raveri}, {Rodriguez-Monroy},
  {Rollins}, {Romer}, {Roodman}, {Rosenfeld}, {Ross}, {Rykoff}, {Samuroff},
  {S{\'a}nchez}, {Sanchez}, {Sanchez}, {Sanchez Cid}, {Scarpine}, {Schubnell},
  {Scolnic}, {Secco}, {Serrano}, {Sevilla-Noarbe}, {Sheldon}, {Shin}, {Smith},
  {Soares-Santos}, {Suchyta}, {Swanson}, {Tabbutt}, {Tarle}, {Thomas}, {To},
  {Troja}, {Troxel}, {Tucker}, {Tutusaus}, {Varga}, {Walker}, {Weaverdyck},
  {Wechsler}, {Weller}, {Yanny}, {Yin}, {Zhang}, {Zuntz}, \& {DES
  Collaboration}}]{DESY3res}
{Abbott}, T.~M.~C., {Aguena}, M., {Alarcon}, A., {et~al.} 2022, \prd, 105,
  023520

\bibitem[{{Akeson} {et~al.}(2019){Akeson}, {Armus}, {Bachelet}, {Bailey},
  {Bartusek}, {Bellini}, {Benford}, {Bennett}, {Bhattacharya}, {Bohlin},
  {et~al.}}]{WFIRSTpap}
{Akeson}, R., {Armus}, L., {Bachelet}, E., {et~al.} 2019, arXiv:1902.05569

\bibitem[{{Albrecht} {et~al.}(2006){Albrecht}, {Bernstein}, {Cahn}, {Freedman},
  {Hewitt}, {Hu}, {Huth}, {Kamionkowski}, {Kolb}, {Knox}, {et~al.}}]{DETFrep}
{Albrecht}, A., {Bernstein}, G., {Cahn}, R., {et~al.} 2006,
  arXiv:astro-ph/0609591

\bibitem[{{Asgari} {et~al.}(2021){Asgari}, {Lin}, {Joachimi}, {Giblin},
  {Heymans}, {Hildebrandt}, {Kannawadi}, {St{\"o}lzner}, {Tr{\"o}ster},
  {et~al.}}]{kids1000}
{Asgari}, M., {Lin}, C.-A., {Joachimi}, B., {et~al.} 2021, \aap, 645, A104

\bibitem[{{Astropy Collaboration} {et~al.}(2018){Astropy Collaboration},
  {Price-Whelan}, {Sip{\H{o}}cz}, {G{\"u}nther}, {Lim}, {Crawford}, {Conseil},
  {Shupe}, {Craig}, {Dencheva}, {et~al.}}]{astropy2}
{Astropy Collaboration}, {Price-Whelan}, A.~M., {Sip{\H{o}}cz}, B.~M., {et~al.}
  2018, \aj, 156, 123

\bibitem[{{Astropy Collaboration} {et~al.}(2013){Astropy Collaboration},
  {Robitaille}, {Tollerud}, {Greenfield}, {Droettboom}, {Bray}, {Aldcroft},
  {Davis}, {Ginsburg}, {Price-Whelan}, {et~al.}}]{astropy1}
{Astropy Collaboration}, {Robitaille}, T.~P., {Tollerud}, E.~J., {et~al.} 2013,
  \aap, 558, A33

\bibitem[{{Barreira} {et~al.}(2018{\natexlab{a}}){Barreira}, {Krause}, \&
  {Schmidt}}]{WLSSCpap}
{Barreira}, A., {Krause}, E., \& {Schmidt}, F. 2018{\natexlab{a}}, \jcap, 2018,
  053

\bibitem[{{Barreira} {et~al.}(2018{\natexlab{b}}){Barreira}, {Krause}, \&
  {Schmidt}}]{2018JCAP...06..015B}
{Barreira}, A., {Krause}, E., \& {Schmidt}, F. 2018{\natexlab{b}}, \jcap, 2018,
  015

\bibitem[{{Barreira} {et~al.}(2019){Barreira}, {Nelson}, {Pillepich},
  {Springel}, {Schmidt}, {Pakmor}, {Hernquist}, \&
  {Vogelsberger}}]{Illustrisbisup}
{Barreira}, A., {Nelson}, D., {Pillepich}, A., {et~al.} 2019, arXiv:1904.02070

\bibitem[{{Bartelmann} \& {Schneider}(2001)}]{BartSchneiRev}
{Bartelmann}, M. \& {Schneider}, P. 2001, \physrep, 340, 291

\bibitem[{{Bernardeau}(1998)}]{BernSLC}
{Bernardeau}, F. 1998, \aap, 338, 375

\bibitem[{{Bernardeau} {et~al.}(2010){Bernardeau}, {Bonvin}, \&
  {Vernizzi}}]{DopplerBernardeau}
{Bernardeau}, F., {Bonvin}, C., \& {Vernizzi}, F. 2010, \prd, 81, 083002

\bibitem[{{Bird} {et~al.}(2012){Bird}, {Viel}, \&
  {Haehnelt}}]{2012MNRAS.420.2551B}
{Bird}, S., {Viel}, M., \& {Haehnelt}, M.~G. 2012, \mnras, 420, 2551

\bibitem[{{Breton} \& {Fleury}(2021)}]{2021A&A...655A..54B}
{Breton}, M.-A. \& {Fleury}, P. 2021, \aap, 655, A54

\bibitem[{{Bridle} \& {King}(2007)}]{IA_NLA}
{Bridle}, S. \& {King}, L. 2007, New Journal of Physics, 9, 444

\bibitem[{{Cooray} \& {Hu}(2002)}]{CoorayHuforder}
{Cooray}, A. \& {Hu}, W. 2002, \apj, 574, 19

\bibitem[{{Cropper} {et~al.}(2012){Cropper}, {Cole}, {James}, {Mellier},
  {Martignac}, {Di Giorgio}, {Paltani}, {Genolet}, {Fourmond}, {Cara},
  {Amiaux}, {Guttridge}, {Walton}, {Thomas}, {Rees}, {Pool}, {Endicott},
  {Holland}, {Gow}, {Murray}, {Duvet}, {Augueres}, {Laureijs}, {Gondoin},
  {Kitching}, {Massey}, \& {Hoekstra}}]{VISpap}
{Cropper}, M., {Cole}, R., {James}, A., {et~al.} 2012, in Society of
  Photo-Optical Instrumentation Engineers (SPIE) Conference Series, Vol. 8442,
  Proc. SPIE, 84420V

\bibitem[{{Cuesta-Lazaro} {et~al.}(2018){Cuesta-Lazaro}, {Quera-Bofarull},
  {Reischke}, \& {Sch{\"a}fer}}]{CuestaLazaroManyCorrs}
{Cuesta-Lazaro}, C., {Quera-Bofarull}, A., {Reischke}, R., \& {Sch{\"a}fer},
  B.~M. 2018, \mnras, 477, 741

\bibitem[{{Deshpande} \& {Kitching}(2020)}]{postlimbrs}
{Deshpande}, A.~C. \& {Kitching}, T.~D. 2020, \prd, 101, 103531

\bibitem[{{Deshpande} \& {Kitching}(2021)}]{DopplerDeshpande}
{Deshpande}, A.~C. \& {Kitching}, T.~D. 2021, \prd, 103, 123510

\bibitem[{{Deshpande} {et~al.}(2020{\natexlab{a}}){Deshpande}, {Kitching},
  {Cardone}, {Taylor}, {Casas}, {Camera}, {Carbone}, {Kilbinger}, {Pettorino},
  {Sakr}, {et~al.}}]{EucRSMB}
{Deshpande}, A.~C., {Kitching}, T.~D., {Cardone}, V.~F., {et~al.}
  2020{\natexlab{a}}, \aap, 636, A95

\bibitem[{{Deshpande} {et~al.}(2020{\natexlab{b}}){Deshpande}, {Taylor}, \&
  {Kitching}}]{kcutrs}
{Deshpande}, A.~C., {Taylor}, P.~L., \& {Kitching}, T.~D. 2020{\natexlab{b}},
  \prd, 102, 083535

\bibitem[{{Desjacques} {et~al.}(2018){Desjacques}, {Jeong}, \&
  {Schmidt}}]{galbiasrev}
{Desjacques}, V., {Jeong}, D., \& {Schmidt}, F. 2018, \physrep, 733, 1

\bibitem[{{Dodelson} {et~al.}(2006){Dodelson}, {Shapiro}, \&
  {White}}]{DodelsonRS}
{Dodelson}, S., {Shapiro}, C., \& {White}, M. 2006, \prd, 73, 023009

\bibitem[{{Duncan} {et~al.}(2022){Duncan}, {Harnois-D{\'e}raps}, {Miller}, \&
  {Langedijk}}]{MBSims}
{Duncan}, C. A.~J., {Harnois-D{\'e}raps}, J., {Miller}, L., \& {Langedijk}, A.
  2022, \mnras, 515, 1130

\bibitem[{{Euclid Collaboration} {et~al.}(2019){Euclid Collaboration},
  {Martinet}, {Schrabback}, {Hoekstra}, {Tewes}, {Herbonnet}, {Schneider},
  {Hernandez-Martin}, {Taylor}, {Brinchmann}, {Carvalho}, {Castellano},
  {Congedo}, {Gillis}, {Jullo}, {K{\"u}mmel}, {Ligori}, {Lilje}, {Padilla},
  {Paris}, {Peacock}, {Pilo}, {Pujol}, {Scott}, \&
  {Toledo-Moreo}}]{2019A&A...627A..59E}
{Euclid Collaboration}, {Martinet}, N., {Schrabback}, T., {et~al.} 2019, \aap,
  627, A59

\bibitem[{{Euclid Collaboration} {et~al.}(2021){Euclid Collaboration},
  {Pocino}, {Tutusaus}, {Castander}, {Fosalba}, {Crocce}, {Porredon}, {Camera},
  {Cardone}, {Casas}, {Kitching}, {Lacasa}, {Martinelli}, {Pourtsidou}, {Sakr},
  {Andreon}, {Auricchio}, {Baccigalupi}, {Balaguera-Antol{\'\i}nez}, {Baldi},
  {Balestra}, {Bardelli}, {Bender}, {Biviano}, {Bodendorf}, {Bonino},
  {Boucaud}, {Bozzo}, {Branchini}, {Brescia}, {Brinchmann}, {Burigana},
  {Cabanac}, {Capobianco}, {Cappi}, {Carvalho}, {Castellano}, {Castignani},
  {Cavuoti}, {Cimatti}, {Cledassou}, {Colodro-Conde}, {Congedo}, {Conselice},
  {Conversi}, {Copin}, {Corcione}, {Costille}, {Coupon}, {Courtois}, {Cropper},
  {Cuby}, {Da Silva}, {de la Torre}, {Di Ferdinando}, {Dubath}, {Duncan},
  {Dupac}, {Dusini}, {Farrens}, {Ferreira}, {Ferrero}, {Finelli}, {Fotopoulou},
  {Frailis}, {Franceschi}, {Galeotta}, {Garilli}, {Gillard}, {Gillis},
  {Giocoli}, {Gozaliasl}, {Graci{\'a}-Carpio}, {Grupp}, {Guzzo}, {Holmes},
  {Hormuth}, {Jahnke}, {Keihanen}, {Kermiche}, {Kiessling}, {Kirkpatrick},
  {Kunz}, {Kurki-Suonio}, {Ligori}, {Lilje}, {Lloro}, {Maino}, {Maiorano},
  {Mansutti}, {Marggraf}, {Martinet}, {Marulli}, {Massey}, {Maurogordato},
  {Medinaceli}, {Mei}, {Meneghetti}, {Benton Metcalf}, {Meylan}, {Moresco},
  {Morin}, {Moscardini}, {Munari}, {Nakajima}, {Neissner}, {Nichol}, {Niemi},
  {Nightingale}, {Padilla}, {Paltani}, {Pasian}, {Patrizii}, {Pedersen},
  {Percival}, {Pettorino}, {Pires}, {Polenta}, {Poncet}, {Popa}, {Potter},
  {Pozzetti}, {Raison}, {Renzi}, {Rhodes}, {Riccio}, {Romelli}, {Roncarelli},
  {Rossetti}, {Saglia}, {S{\'a}nchez}, {Sapone}, {Scaramella}, {Schneider},
  {Scottez}, {Secroun}, {Seidel}, {Serrano}, {Sirignano}, {Sirri}, {Stanco},
  {Sureau}, {Taylor}, {Tenti}, {Tereno}, {Teyssier}, {Toledo-Moreo},
  {Tramacere}, {Valentijn}, {Valenziano}, {Valiviita}, {Vassallo}, {Viel},
  {Wang}, {Welikala}, {Whittaker}, {Zacchei}, {Zamorani}, {Zoubian}, \&
  {Zucca}}]{2021A&A...655A..44E}
{Euclid Collaboration}, {Pocino}, A., {Tutusaus}, I., {et~al.} 2021, \aap, 655,
  A44

\bibitem[{{Euclid Collaboration: Blanchard} {et~al.}(2020){Euclid
  Collaboration: Blanchard}, {Camera}, {Carbone}, {Cardone}, {Casas}, {Clesse},
  {Ili{\'c}}, {Kilbinger}, {Kitching}, {et~al.}}]{ISTforecast}
{Euclid Collaboration: Blanchard}, A., {Camera}, S., {Carbone}, C., {et~al.}
  2020, \aap, 642, A191

\bibitem[{{Euclid Collaboration: Bretonni{\`e}re} {et~al.}(2022){Euclid
  Collaboration: Bretonni{\`e}re}, {Huertas-Company}, {Boucaud}, {Lanusse},
  {Jullo}, {Merlin}, {Tuccillo}, {Castellano}, {Brinchmann}, {Conselice},
  {Dole}, {Cabanac}, {Courtois}, {Castander}, {Duc}, {Fosalba}, {Guinet},
  {Kruk}, {Kuchner}, {Serrano}, {Soubrie}, {Tramacere}, {Wang}, {Amara},
  {Auricchio}, {Bender}, {Bodendorf}, {Bonino}, {Branchini}, {Brau-Nogue},
  {Brescia}, {Capobianco}, {Carbone}, {Carretero}, {Cavuoti}, {Cimatti},
  {Cledassou}, {Congedo}, {Conversi}, {Copin}, {Corcione}, {Costille},
  {Cropper}, {Da Silva}, {Degaudenzi}, {Douspis}, {Dubath}, {Duncan}, {Dupac},
  {Dusini}, {Farrens}, {Ferriol}, {Frailis}, {Franceschi}, {Fumana}, {Garilli},
  {Gillard}, {Gillis}, {Giocoli}, {Grazian}, {Grupp}, {Haugan}, {Holmes},
  {Hormuth}, {Hudelot}, {Jahnke}, {Kermiche}, {Kiessling}, {Kilbinger},
  {Kitching}, {Kohley}, {K{\"u}mmel}, {Kunz}, {Kurki-Suonio}, {Ligori},
  {Lilje}, {Lloro}, {Maiorano}, {Mansutti}, {Marggraf}, {Markovic}, {Marulli},
  {Massey}, {Maurogordato}, {Melchior}, {Meneghetti}, {Meylan}, {Moresco},
  {Morin}, {Moscardini}, {Munari}, {Nakajima}, {Niemi}, {Padilla}, {Paltani},
  {Pasian}, {Pedersen}, {Pettorino}, {Pires}, {Poncet}, {Popa}, {Pozzetti},
  {Raison}, {Rebolo}, {Rhodes}, {Roncarelli}, {Rossetti}, {Saglia},
  {Schneider}, {Secroun}, {Seidel}, {Sirignano}, {Sirri}, {Stanco}, {Starck},
  {Tallada-Cresp{\'\i}}, {Taylor}, {Tereno}, {Toledo-Moreo}, {Torradeflot},
  {Valentijn}, {Valenziano}, {Wang}, {Welikala}, {Weller}, {Zamorani},
  {Zoubian}, {Baldi}, {Bardelli}, {Camera}, {Farinelli}, {Medinaceli}, {Mei},
  {Polenta}, {Romelli}, {Tenti}, {Vassallo}, {Zacchei}, {Zucca}, {Baccigalupi},
  {Balaguera-Antol{\'\i}nez}, {Biviano}, {Borgani}, {Bozzo}, {Burigana},
  {Cappi}, {Carvalho}, {Casas}, {Castignani}, {Colodro-Conde}, {Coupon}, {de la
  Torre}, {Fabricius}, {Farina}, {Ferreira}, {Flose-Reimberg}, {Fotopoulou},
  {Galeotta}, {Ganga}, {Garcia-Bellido}, {Gaztanaga}, {Gozaliasl}, {Hook},
  {Joachimi}, {Kansal}, {Kashlinsky}, {Keihanen}, {Kirkpatrick}, {Lindholm},
  {Mainetti}, {Maino}, {Maoli}, {Martinelli}, {Martinet}, {McCracken},
  {Metcalf}, {Morgante}, {Morisset}, {Nightingale}, {Nucita}, {Patrizii},
  {Potter}, {Renzi}, {Riccio}, {S{\'a}nchez}, {Sapone}, {Schirmer},
  {Schultheis}, {Scottez}, {Sefusatti}, {Teyssier}, {Tutusaus}, {Valiviita},
  {Viel}, {Whittaker}, \& {Knapen}}]{EuclidFlagshipGalsize}
{Euclid Collaboration: Bretonni{\`e}re}, H., {Huertas-Company}, M., {Boucaud},
  A., {et~al.} 2022, \aap, 657, A90

\bibitem[{{Euclid Collaboration: Lepori} {et~al.}(2022){Euclid Collaboration:
  Lepori}, {Tutusaus}, {Viglione}, {Bonvin}, {Camera}, {Castander}, {Durrer},
  {Fosalba}, {Jelic-Cizmek}, {Kunz}, {Adamek}, {Casas}, {Martinelli}, {Sakr},
  {Sapone}, {Amara}, {Auricchio}, {Bodendorf}, {Bonino}, {Branchini},
  {Brescia}, {Brinchmann}, {Capobianco}, {Carbone}, {Carretero}, {Castellano},
  {Cavuoti}, {Cimatti}, {Cledassou}, {Congedo}, {Conselice}, {Conversi},
  {Copin}, {Corcione}, {Courbin}, {Da Silva}, {Degaudenzi}, {Douspis},
  {Dubath}, {Dupac}, {Dusini}, {Ealet}, {Farrens}, {Ferriol}, {Franceschi},
  {Fumana}, {Garilli}, {Gillard}, {Gillis}, {Giocoli}, {Grazian}, {Grupp},
  {Guzzo}, {Haugan}, {Holmes}, {Hormuth}, {Hudelot}, {Jahnke}, {Kermiche},
  {Kiessling}, {Kilbinger}, {Kitching}, {K{\"u}mmel}, {Kurki-Suonio}, {Ligori},
  {Lilje}, {Lloro}, {Mansutti}, {Marggraf}, {Markovic}, {Marulli}, {Massey},
  {Maurogordato}, {Melchior}, {Meneghetti}, {Merlin}, {Meylan}, {Moresco},
  {Moscardini}, {Munari}, {Nakajima}, {Niemi}, {Padilla}, {Paltani}, {Pasian},
  {Pedersen}, {Percival}, {Pettorino}, {Pires}, {Poncet}, {Popa}, {Pozzetti},
  {Raison}, {Rhodes}, {Roncarelli}, {Rossetti}, {Saglia}, {Schneider},
  {Secroun}, {Seidel}, {Serrano}, {Sirignano}, {Sirri}, {Stanco}, {Starck},
  {Tallada-Cresp{\'\i}}, {Taylor}, {Tereno}, {Toledo-Moreo}, {Torradeflot},
  {Valentijn}, {Valenziano}, {Wang}, {Weller}, {Zamorani}, {Zoubian},
  {Andreon}, {Bardelli}, {Fabbian}, {Graci{\'a}-Carpio}, {Maino}, {Medinaceli},
  {Mei}, {Renzi}, {Romelli}, {Sureau}, {Vassallo}, {Zacchei}, {Zucca},
  {Baccigalupi}, {Balaguera-Antol{\'\i}nez}, {Bernardeau}, {Biviano},
  {Blanchard}, {Bolzonella}, {Borgani}, {Bozzo}, {Burigana}, {Cabanac},
  {Cappi}, {Carvalho}, {Castignani}, {Colodro-Conde}, {Coupon}, {Courtois},
  {Cuby}, {Davini}, {de la Torre}, {Di Ferdinando}, {Farina}, {Ferreira},
  {Finelli}, {Galeotta}, {Ganga}, {Garcia-Bellido}, {Gaztanaga}, {Gozaliasl},
  {Hook}, {Ili{\'c}}, {Joachimi}, {Kansal}, {Keihanen}, {Kirkpatrick},
  {Lindholm}, {Mainetti}, {Maoli}, {Martinet}, {Maturi}, {Metcalf}, {Monaco},
  {Morgante}, {Nightingale}, {Nucita}, {Patrizii}, {Popa}, {Potter}, {Riccio},
  {S{\'a}nchez}, {Schirmer}, {Schultheis}, {Scottez}, {Sefusatti}, {Tramacere},
  {Valiviita}, {Viel}, \& {Hildebrandt}}]{EuclidMagGCWL}
{Euclid Collaboration: Lepori}, F., {Tutusaus}, I., {Viglione}, C., {et~al.}
  2022, \aap, 662, A93

\bibitem[{Fleury {et~al.}(2017)Fleury, Larena, \& Uzan}]{finitebeamcoe}
Fleury, P., Larena, J., \& Uzan, J.-P. 2017, \prl, 119, 191101

\bibitem[{{Fleury} {et~al.}(2019){Fleury}, {Larena}, \&
  {Uzan}}]{2019PhRvD..99b3525F}
{Fleury}, P., {Larena}, J., \& {Uzan}, J.-P. 2019, \prd, 99, 023525

\bibitem[{{Fortuna} {et~al.}(2021){Fortuna}, {Hoekstra}, {Johnston}, {Vakili},
  {Kannawadi}, {Georgiou}, {Joachimi}, {Wright}, {Asgari}, {Bilicki},
  {Heymans}, {Hildebrandt}, {Kuijken}, \& {Von
  Wietersheim-Kramsta}}]{FortunaIA}
{Fortuna}, M.~C., {Hoekstra}, H., {Johnston}, H., {et~al.} 2021, \aap, 654, A76

\bibitem[{{Gil-Mar{\'\i}n} {et~al.}(2012){Gil-Mar{\'\i}n}, {Wagner},
  {Fragkoudi}, {Jimenez}, \& {Verde}}]{Gilmarinbispec}
{Gil-Mar{\'\i}n}, H., {Wagner}, C., {Fragkoudi}, F., {Jimenez}, R., \& {Verde},
  L. 2012, \jcap, 2012, 047

\bibitem[{{Hall}(2020)}]{LUHall}
{Hall}, A. 2020, \prd, 101, 043519

\bibitem[{{Hall} \& {Taylor}(2022)}]{HT22}
{Hall}, A. \& {Taylor}, A. 2022, \prd, 105, 123527

\bibitem[{{Hamana} {et~al.}(2002){Hamana}, {Colombi}, {Thion}, {Devriendt},
  {Mellier}, \& {Bernardeau}}]{HamanaSLC}
{Hamana}, T., {Colombi}, S.~T., {Thion}, A., {et~al.} 2002, \mnras, 330, 365

\bibitem[{{Hartlap} {et~al.}(2011){Hartlap}, {Hilbert}, {Schneider}, \&
  {Hildebrandt}}]{SourceObs}
{Hartlap}, J., {Hilbert}, S., {Schneider}, P., \& {Hildebrandt}, H. 2011, \aap,
  528, A51

\bibitem[{{Heydenreich} {et~al.}(2020){Heydenreich}, {Schneider},
  {Hildebrandt}, {Asgari}, {Heymans}, {Joachimi}, {Kuijken}, {Lin},
  {Tr{\"o}ster}, \& {van den Busch}}]{heydenreichspatdepth}
{Heydenreich}, S., {Schneider}, P., {Hildebrandt}, H., {et~al.} 2020, \aap,
  634, A104

\bibitem[{{Hikage} {et~al.}(2019){Hikage}, {Oguri}, {Hamana}, {More},
  {Mandelbaum}, {Takada}, {K{\"o}hlinger}, {Miyatake}, {Nishizawa}, {Aihara},
  {Armstrong}, {Bosch}, {Coupon}, {Ducout}, {Ho}, {Hsieh}, {Komiyama},
  {Lanusse}, {Leauthaud}, {Lupton}, {Medezinski}, {Mineo}, {Miyama},
  {Miyazaki}, {Murata}, {Murayama}, {Shirasaki}, {Sif{\'o}n}, {Simet},
  {Speagle}, {Spergel}, {Strauss}, {Sugiyama}, {Tanaka}, {Utsumi}, {Wang}, \&
  {Yamada}}]{HSCpowerspec}
{Hikage}, C., {Oguri}, M., {Hamana}, T., {et~al.} 2019, \pasj, 71, 43

\bibitem[{{Hu}(1999)}]{1999ApJ...522L..21H}
{Hu}, W. 1999, \apjl, 522, L21

\bibitem[{{Hu} \& {Kravtsov}(2003)}]{SSCorig}
{Hu}, W. \& {Kravtsov}, A.~V. 2003, \apj, 584, 702

\bibitem[{{Hui} {et~al.}(2007){Hui}, {Gazta{\~n}aga}, \&
  {Loverde}}]{MBcorssource}
{Hui}, L., {Gazta{\~n}aga}, E., \& {Loverde}, M. 2007, \prd, 76, 103502

\bibitem[{{Joachimi} {et~al.}(2015){Joachimi}, {Cacciato}, {Kitching},
  {Leonard}, {Mandelbaum}, {Sch{\"a}fer}, {Sif{\'o}n}, {Hoekstra}, {Kiessling},
  {Kirk}, {et~al.}}]{IA1}
{Joachimi}, B., {Cacciato}, M., {Kitching}, T., {et~al.} 2015, \ssr, 193, 1

\bibitem[{{Kaiser}(1992)}]{Kaiser92}
{Kaiser}, N. 1992, \apj, 388, 272

\bibitem[{{Kaiser}(1998)}]{KaiserLimber}
{Kaiser}, N. 1998, \apj, 498, 26

\bibitem[{{Kiessling} {et~al.}(2015){Kiessling}, {Cacciato}, {Joachimi},
  {Kirk}, {Kitching}, {Leonard}, {Mandelbaum}, {Sch{\"a}fer}, {Sif{\'o}n},
  {Brown}, {et~al.}}]{IA3}
{Kiessling}, A., {Cacciato}, M., {Joachimi}, B., {et~al.} 2015, \ssr, 193, 67

\bibitem[{Kilbinger(2015)}]{Kilbinger15}
Kilbinger, M. 2015, Reports on Progress in Physics, 78, 086901

\bibitem[{{Kirk} {et~al.}(2015){Kirk}, {Brown}, {Hoekstra}, {Joachimi},
  {Kitching}, {Mandelbaum}, {Sif{\'o}n}, {Cacciato}, {Choi}, {Kiessling},
  {et~al.}}]{IA2}
{Kirk}, D., {Brown}, M., {Hoekstra}, H., {et~al.} 2015, \ssr, 193, 139

\bibitem[{{Kitching} {et~al.}(2021){Kitching}, {Deshpande}, \&
  {Taylor}}]{ShapeMeas21}
{Kitching}, T., {Deshpande}, A., \& {Taylor}, P. 2021, The Open Journal of
  Astrophysics, 4, 17

\bibitem[{{Kitching} {et~al.}(2017){Kitching}, {Alsing}, {Heavens}, {Jimenez},
  {McEwen}, \& {Verde}}]{limitsofshear17}
{Kitching}, T.~D., {Alsing}, J., {Heavens}, A.~F., {et~al.} 2017, \mnras, 469,
  2737

\bibitem[{{Kitching} \& {Deshpande}(2022)}]{ShapeMeas22}
{Kitching}, T.~D. \& {Deshpande}, A.~C. 2022, The Open Journal of Astrophysics,
  5, 6

\bibitem[{{Kitching} {et~al.}(2020){Kitching}, {Deshpande}, \&
  {Taylor}}]{ShapeMeas20}
{Kitching}, T.~D., {Deshpande}, A.~C., \& {Taylor}, P.~L. 2020, The Open
  Journal of Astrophysics, 3, 14

\bibitem[{{Kitching} \& {Heavens}(2017)}]{Unequaltimecorr}
{Kitching}, T.~D. \& {Heavens}, A.~F. 2017, \prd, 95, 063522

\bibitem[{{Kitching} {et~al.}(2019){Kitching}, {Paykari}, {Hoekstra}, \&
  {Cropper}}]{ShapeMeas19}
{Kitching}, T.~D., {Paykari}, P., {Hoekstra}, H., \& {Cropper}, M. 2019, The
  Open Journal of Astrophysics, 2, 5

\bibitem[{{Kitching} {et~al.}(2008){Kitching}, {Taylor}, \&
  {Heavens}}]{2008MNRAS.389..173K}
{Kitching}, T.~D., {Taylor}, A.~N., \& {Heavens}, A.~F. 2008, \mnras, 389, 173

\bibitem[{{Krause} \& {Hirata}(2010)}]{KrauseHirataRS}
{Krause}, E. \& {Hirata}, C. 2010, \aap, 523, A28

\bibitem[{{Lacasa} \& {Grain}(2019)}]{fastssc}
{Lacasa}, F. \& {Grain}, J. 2019, \aap, 624, A61

\bibitem[{{Laureijs} {et~al.}(2011){Laureijs}, {Amiaux}, {Arduini},
  {Augu{\`e}res}, {Brinchmann}, {Cole}, {Cropper}, {Dabin}, {Duvet}, {Ealet},
  {et~al.}}]{EuclidRB}
{Laureijs}, R., {Amiaux}, J., {Arduini}, S., {et~al.} 2011, arXiv:1110.3193

\bibitem[{{Lavaux} \& {Hudson}(2011)}]{2M++}
{Lavaux}, G. \& {Hudson}, M.~J. 2011, \mnras, 416, 2840

\bibitem[{{Lesgourgues} \& {Tram}(2014)}]{curvprefacpap}
{Lesgourgues}, J. \& {Tram}, T. 2014, \jcap, 2014, 032

\bibitem[{{Lewis} {et~al.}(2000){Lewis}, {Challinor}, \& {Lasenby}}]{cambpap}
{Lewis}, A., {Challinor}, A., \& {Lasenby}, A. 2000, \apj, 538, 473

\bibitem[{{Limber}(1953)}]{Limberorig}
{Limber}, D.~N. 1953, \apj, 117, 134

\bibitem[{{Lin} {et~al.}(2019){Lin}, {Harnois-D{\'e}raps}, {Eifler},
  {Pospisil}, {Mandelbaum}, {Lee}, \& {Singh}}]{LSSTnonGauss}
{Lin}, C.-H., {Harnois-D{\'e}raps}, J., {Eifler}, T., {et~al.} 2019,
  arXiv:1905.03779

\bibitem[{{Loureiro} {et~al.}(2021){Loureiro}, {Whittaker}, {Spurio Mancini},
  {Joachimi}, {Cuceu}, {Asgari}, {St{\"o}lzner}, {Tr{\"o}ster}, {Wright},
  {Bilicki}, {et~al.}}]{vardepthfix}
{Loureiro}, A., {Whittaker}, L., {Spurio Mancini}, A., {et~al.} 2021,
  arXiv:2110.06947

\bibitem[{{LoVerde} \& {Afshordi}(2008)}]{ExtendedLimber}
{LoVerde}, M. \& {Afshordi}, N. 2008, \prd, 78, 123506

\bibitem[{{LSST Science Collaboration} {et~al.}(2009){LSST Science
  Collaboration}, {Abell}, {Allison}, {Anderson}, {Andrew}, {Angel}, {Armus},
  {Arnett}, {Asztalos}, {Axelrod}, {et~al.}}]{LSSTpap}
{LSST Science Collaboration}, {Abell}, P.~A., {Allison}, J., {et~al.} 2009,
  arXiv:0912.0201

\bibitem[{{Martinelli} {et~al.}(2021){Martinelli}, {Tutusaus}, {Archidiacono},
  {Camera}, {Cardone}, {Clesse}, {Casas}, {Casarini}, {Mota}, {Hoekstra},
  {Carbone}, {Ili{\'c}}, {Kitching}, {Pettorino}, {Pourtsidou}, {Sakr},
  {Sapone}, {Auricchio}, {Balestra}, {Boucaud}, {Branchini}, {Brescia},
  {Capobianco}, {Carretero}, {Castellano}, {Cavuoti}, {Cimatti}, {Cledassou},
  {Congedo}, {Conselice}, {Conversi}, {Corcione}, {Costille}, {Douspis},
  {Dubath}, {Dusini}, {Fabbian}, {Fosalba}, {Frailis}, {Franceschi}, {Gillis},
  {Giocoli}, {Grupp}, {Guzzo}, {Holmes}, {Hormuth}, {Jahnke}, {Kermiche},
  {Kiessling}, {Kilbinger}, {Kunz}, {Kurki-Suonio}, {Ligori}, {Lilje}, {Lloro},
  {Maiorano}, {Marggraf}, {Markovic}, {Massey}, {Meneghetti}, {Meylan},
  {Morin}, {Moscardini}, {Niemi}, {Padilla}, {Paltani}, {Pasian}, {Pedersen},
  {Pires}, {Polenta}, {Poncet}, {Popa}, {Raison}, {Rhodes}, {Roncarelli},
  {Rossetti}, {Saglia}, {Schneider}, {Secroun}, {Serrano}, {Sirignano},
  {Sirri}, {Starck}, {Sureau}, {Taylor}, {Tereno}, {Toledo-Moreo}, {Valentijn},
  {Valenziano}, {Vassallo}, {Wang}, {Welikala}, {Zacchei}, \&
  {Zoubian}}]{2021A&A...649A.100M}
{Martinelli}, M., {Tutusaus}, I., {Archidiacono}, M., {et~al.} 2021, \aap, 649,
  A100

\bibitem[{{Munshi} {et~al.}(2008){Munshi}, {Valageas}, {van Waerbeke}, \&
  {Heavens}}]{Munshirev}
{Munshi}, D., {Valageas}, P., {van Waerbeke}, L., \& {Heavens}, A. 2008,
  \physrep, 462, 67

\bibitem[{{Planck Collaboration} {et~al.}(2018){Planck Collaboration},
  {Aghanim}, {Akrami}, {Ashdown}, {Aumont}, {Baccigalupi}, {Ballardini},
  {Banday}, {Barreiro}, {Bartolo}, {et~al.}}]{Planck18}
{Planck Collaboration}, {Aghanim}, N., {Akrami}, Y., {et~al.} 2018,
  arXiv:1807.06209

\bibitem[{{Potter} {et~al.}(2017){Potter}, {Stadel}, \&
  {Teyssier}}]{Flagshippaper}
{Potter}, D., {Stadel}, J., \& {Teyssier}, R. 2017, Computational Astrophysics
  and Cosmology, 4, 2

\bibitem[{{Reischke} {et~al.}(2019){Reischke}, {Sch{\"a}fer}, {Bolejko},
  {Lewis}, \& {Lautsch}}]{LUOrig}
{Reischke}, R., {Sch{\"a}fer}, B.~M., {Bolejko}, K., {Lewis}, G.~F., \&
  {Lautsch}, M. 2019, \mnras, 486, 5061

\bibitem[{{Schmidt} {et~al.}(2009){Schmidt}, {Rozo}, {Dodelson}, {Hui}, \&
  {Sheldon}}]{RSMBcombpap}
{Schmidt}, F., {Rozo}, E., {Dodelson}, S., {Hui}, L., \& {Sheldon}, E. 2009,
  \apj, 702, 593

\bibitem[{{Schneider} \& {Er}(2008)}]{flexionpap}
{Schneider}, P. \& {Er}, X. 2008, \aap, 485, 363

\bibitem[{{Schneider} {et~al.}(2002){Schneider}, {van Waerbeke}, \&
  {Mellier}}]{Bmodespap}
{Schneider}, P., {van Waerbeke}, L., \& {Mellier}, Y. 2002, \aap, 389, 729

\bibitem[{{Scoccimarro} \& {Couchman}(2001)}]{Scoccimarro01}
{Scoccimarro}, R. \& {Couchman}, H. 2001, \mnras, 325, 1312

\bibitem[{{Seitz} \& {Schneider}(1997)}]{1997A&A...318..687S}
{Seitz}, C. \& {Schneider}, P. 1997, \aap, 318, 687

\bibitem[{{Semboloni} {et~al.}(2013){Semboloni}, {Hoekstra}, \&
  {Schaye}}]{Owlsbisup}
{Semboloni}, E., {Hoekstra}, H., \& {Schaye}, J. 2013, \mnras, 434, 148

\bibitem[{Shapiro(2009)}]{Shapiro09}
Shapiro, C. 2009, \apj, 696, 775

\bibitem[{{Shapiro} \& {Cooray}(2006)}]{ShapCooray}
{Shapiro}, C. \& {Cooray}, A. 2006, \jcap, 2006, 007

\bibitem[{{Simpson} {et~al.}(2010){Simpson}, {Peacock}, \&
  {Heavens}}]{lensingbylambda}
{Simpson}, F., {Peacock}, J.~A., \& {Heavens}, A.~F. 2010, \mnras, 402, 2009

\bibitem[{{Spurio Mancini} {et~al.}(2022){Spurio Mancini}, {Piras}, {Alsing},
  {Joachimi}, \& {Hobson}}]{cosmopower}
{Spurio Mancini}, A., {Piras}, D., {Alsing}, J., {Joachimi}, B., \& {Hobson},
  M.~P. 2022, \mnras, 511, 1771

\bibitem[{{Takada} \& {Hu}(2013)}]{HuSSC}
{Takada}, M. \& {Hu}, W. 2013, \prd, 87, 123504

\bibitem[{{Takahashi} {et~al.}(2020){Takahashi}, {Nishimichi}, {Namikawa},
  {Taruya}, {Kayo}, {Osato}, {Kobayashi}, \& {Shirasaki}}]{bihalofit}
{Takahashi}, R., {Nishimichi}, T., {Namikawa}, T., {et~al.} 2020, \apj, 895,
  113

\bibitem[{{Takahashi} {et~al.}(2012){Takahashi}, {Sato}, {Nishimichi},
  {Taruya}, \& {Oguri}}]{Takahashi12}
{Takahashi}, R., {Sato}, M., {Nishimichi}, T., {Taruya}, A., \& {Oguri}, M.
  2012, \apj, 761, 152

\bibitem[{{Taylor} {et~al.}(2007){Taylor}, {Kitching}, {Bacon}, \&
  {Heavens}}]{biaspap}
{Taylor}, A., {Kitching}, T., {Bacon}, D., \& {Heavens}, A. 2007, \mnras, 374,
  1377

\bibitem[{{Taylor} {et~al.}(2018{\natexlab{a}}){Taylor}, {Bernardeau}, \&
  {Kitching}}]{kcutpap}
{Taylor}, P.~L., {Bernardeau}, F., \& {Kitching}, T.~D. 2018{\natexlab{a}},
  \prd, 98, 083514

\bibitem[{{Taylor} {et~al.}(2019{\natexlab{a}}){Taylor}, {Kitching}, {Alsing},
  {Wandelt}, {Feeney}, \& {McEwen}}]{TaylorNG}
{Taylor}, P.~L., {Kitching}, T.~D., {Alsing}, J., {et~al.} 2019{\natexlab{a}},
  \prd, 100, 023519

\bibitem[{{Taylor} {et~al.}(2019{\natexlab{b}}){Taylor}, {Kitching}, {Alsing},
  {Wandelt}, {Feeney}, \& {McEwen}}]{2019PhRvD.100b3519T}
{Taylor}, P.~L., {Kitching}, T.~D., {Alsing}, J., {et~al.} 2019{\natexlab{b}},
  \prd, 100, 023519

\bibitem[{{Taylor} {et~al.}(2018{\natexlab{b}}){Taylor}, {Kitching}, {McEwen},
  \& {Tram}}]{Spatflatpap}
{Taylor}, P.~L., {Kitching}, T.~D., {McEwen}, J.~D., \& {Tram}, T.
  2018{\natexlab{b}}, \prd, 98, 023522

\bibitem[{Tegmark {et~al.}(2015)Tegmark, Taylor, \& Heavens}]{Tegmark97}
Tegmark, M., Taylor, A., \& Heavens, A. 2015, \apj, 78, 086901

\bibitem[{{Turner} {et~al.}(1984){Turner}, {Ostriker}, \& {Gott}}]{MBorig}
{Turner}, E.~L., {Ostriker}, J.~P., \& {Gott}, III, J.~R. 1984, \apj, 284, 1

\bibitem[{{Upham} {et~al.}(2021){Upham}, {Brown}, \& {Whittaker}}]{UphamGauss}
{Upham}, R.~E., {Brown}, M.~L., \& {Whittaker}, L. 2021, \mnras, 503, 1999

\bibitem[{{Upham} {et~al.}(2022){Upham}, {Brown}, {Whittaker}, {Amara},
  {Auricchio}, {Bonino}, {Branchini}, {Brescia}, {Brinchmann}, {Capobianco},
  {Carbone}, {Carretero}, {Castellano}, {Cavuoti}, {Cimatti}, {Cledassou},
  {Congedo}, {Conversi}, {Copin}, {Corcione}, {Cropper}, {Da Silva},
  {Degaudenzi}, {Douspis}, {Dubath}, {Duncan}, {Dupac}, {Dusini}, {Ealet},
  {Farrens}, {Ferriol}, {Fosalba}, {Frailis}, {Franceschi}, {Fumana},
  {Garilli}, {Gillis}, {Giocoli}, {Grupp}, {Haugan}, {Hoekstra}, {Holmes},
  {Hormuth}, {Hornstrup}, {Jahnke}, {Kermiche}, {Kiessling}, {Kilbinger},
  {Kitching}, {K{\"u}mmel}, {Kunz}, {Kurki-Suonio}, {Ligori}, {Lilje}, {Lloro},
  {Marggraf}, {Markovic}, {Marulli}, {Meneghetti}, {Meylan}, {Moresco},
  {Moscardini}, {Munari}, {Niemi}, {Padilla}, {Paltani}, {Pasian}, {Pedersen},
  {Pettorino}, {Pires}, {Poncet}, {Popa}, {Raison}, {Rhodes}, {Rossetti},
  {Saglia}, {Sartoris}, {Schneider}, {Secroun}, {Seidel}, {Sirignano}, {Sirri},
  {Stanco}, {Starck}, {Tallada-Cresp{\'\i}}, {Tavagnacco}, {Taylor}, {Tereno},
  {Toledo-Moreo}, {Torradeflot}, {Valenziano}, {Wang}, {Zamorani}, {Zoubian},
  {Andreon}, {Baldi}, {Camera}, {Cardone}, {Fabbian}, {Polenta}, {Renzi},
  {Joachimi}, {Hall}, {Loureiro}, \& {Sellentin}}]{UphamNonGauss}
{Upham}, R.~E., {Brown}, M.~L., {Whittaker}, L., {et~al.} 2022, \aap, 660, A114

\bibitem[{{Viola} {et~al.}(2014){Viola}, {Kitching}, \&
  {Joachimi}}]{2014MNRAS.439.1909V}
{Viola}, M., {Kitching}, T.~D., \& {Joachimi}, B. 2014, \mnras, 439, 1909

\bibitem[{{Yu} {et~al.}(2015){Yu}, {Zhang}, {Lin}, \& {Cui}}]{YuSLC}
{Yu}, Y., {Zhang}, P., {Lin}, W., \& {Cui}, W. 2015, \apj, 803, 46

\end{thebibliography}

\end{document}